\documentclass[aps,pre,twocolumn,longbibliography,showpacs]{revtex4-1}
\usepackage{amsfonts,amssymb,amsmath,latexsym,epsfig}
\usepackage{xcolor}
\usepackage{graphics}
\usepackage{epsfig}
\usepackage{color}

\newcommand{\be}{\begin{equation}}
\newcommand{\ee}{\end{equation}}
\newcommand{\bea}{\begin{eqnarray}}
\newcommand{\eea}{\end{eqnarray}}

\begin{document}

\title{Coherent oscillations in balanced neural networks driven by endogenous fluctuations
}

\author{Matteo di Volo$^{1}$}
\email{matteo.divolo@cyu.fr}
\author{Marco Segneri$^{1}$}
\email{marco.segneri@cyu.fr}
\author{Denis Goldobin$^{2,3}$}
\email{denis.goldobin@gmail.com}
\author{Antonio Politi$^{4}$}
\email{a.politi@abdn.ac.uk}
\author{Alessandro Torcini$^{1,5,6}$}
\email{alessandro.torcini@cyu.fr}
\affiliation{ $^{1}$CY Cergy Paris Universit\'e, CNRS, Laboratoire de Physique Th\'eorique et Mod\'elisation, UMR 8089,
95302 Cergy-Pontoise , France \\
$^{2}$ Institute of Continuous Media Mechanics, Ural Branch of RAS, Acad. Korolevstreet 1, 614013 Perm, Russia\\
$^{3}$ Department of Theoretical Physics, Perm State University, Bukirev street 15,614099 Perm, Russia\\
$^{4}$
Institute for Pure and Applied Mathematics and Department of Physics (SUPA), Old Aberdeen, Aberdeen AB24 3UE, United Kingdom \\
$^{5}$ CNR - Consiglio Nazionale delle Ricerche - Istituto dei Sistemi Complessi, via Madonna del Piano 10, 50019 Sesto Fiorentino, Italy \\
$^{6}$ INFN Sezione di Firenze, Via Sansone 1, I-50019 Sesto Fiorentino, Florence, Italy}

\date{\today}

\begin{abstract}
We present a detailed analysis of the dynamical regimes observed in a balanced network 
of identical Quadratic Integrate-and-Fire (QIF) neurons with a sparse connectivity 
for homogeneous and heterogeneous in-degree distribution.
Depending on the parameter values, either an asynchronous regime or periodic oscillations spontaneously
emerge.
Numerical simulations are compared with a mean field model based on a self-consistent Fokker-Planck equation (FPE).
The FPE reproduces quite well the asynchronous dynamics in the 
homogeneous case by either assuming a Poissonian or renewal distribution for the 
incoming spike trains. 
An exact self consistent solution for the mean firing rate obtained in the limit of infinite in-degree allows identifying
balanced regimes that can be either mean- or fluctuation-driven. 
A low-dimensional reduction of the FPE in terms of circular cumulants is also considered.
Two cumulants suffice to reproduce the transition scenario observed in the network.
The emergence  of periodic collective oscillations is well captured both in the homogeneous and heterogeneous 
set-ups by the mean field models upon tuning either the connectivity, or the input DC current. 
In the heterogeneous situation we analyze also the role of structural heterogeneity.
\end{abstract}


\maketitle

{
\bf
The balance of excitation and inhibition represents a crucial
aspect of brain dynamics explaining the highly irregular
fluctuations observed in several parts of the brain.
The identification of macroscopic phases emerging spontaneously in balanced neural
networks is particularly relevant in neuroscience since classifying them and
establishing their robustness (generality) can help to understand and control brain functions.
Focusing on pulse coupled Quadratic Integrate-and-Fire neurons 
we illustrate and describe in a quantitative way, the asynchronous dynamics and the emergence of collective oscillations.
Our main assumption is that the spontaneous current fluctuations emerging in the
network due to the sparseness of the connections can be assimilated to (white) noise whose amplitude 
is determined self-consistently.
This way the dimensionality of the collective dynamics is ``reduced" to that of a nonlinear
Fokker-Planck equation, a quite effective reduction to few degrees of freedom is also implemented.
}

\section{Introduction}

The emergence of collective oscillations (COs) in complex systems
has been extensively studied in the last 50 years
from an experimental as well as a theoretical point of view \cite{pikovsky2015}. 
Statistical mechanics and nonlinear dynamics
approaches have been employed to describe
networks of heterogeneous oscillators \cite{winfree,kuramoto2012,hong2007, crawford1994,strogatz2000,barre2016}.
Furthermore, exact analytic reduction methodologies have been developed, which allow
passing from infinite dimensional dynamics to few macroscopic variables in some
homogeneous \cite{watanabe1994} and heterogeneous \cite{ott2008} globally coupled networks of phase oscillators.

In the last years these reduction techniques have been
extended to globally coupled spiking neural networks 
either heterogeneous \cite{luke2013,montbrio2015}
or homogeneous \cite{laing2018}, thus opening new perspectives for 
the study of large ensembles of spiking neurons and for the understanding
of the mechanisms underlying brain rhythms \cite{buzsaki2006}.
The reduction methodologies have been usually limited to globally coupled systems
in absence of either noise or spatial disorder; only recently they
have been extended to  noisy systems \cite{tyulkina2018,ratas2019} and
sparse neural networks \cite{matteo,goldobin2021}. 

Cortical neurons are subject to a continuous barrage  from thousands of pre-synaptic neurons, 
a stimulation which is intuitively expected to induce an almost constant depolarization of the neurons and,
thereby, a regular firing activity. However, cortical neurons fire irregularly at a low rate~\cite{softky1992}.
This apparent contradiction can be solved in the so-called balanced network, 
where the current is affected by strong statistical fluctuations as a result of the 
approximately equal strength of excitatory and inhibitory synaptic drives~\cite{bal1}. 
Balanced asynchronous irregular dynamics has
been experimentally reported both {\it in vivo} and {\it in vitro} \cite{shu2003,haider2006,barral2016}.
A balance of excitation and inhibition appears to be
crucial also for the emergence of cortical oscillations and in brain rhythms \cite{okun2008,isaacson2011,le2016}.
 
Stationary irregular activity may manifest itself either in the form of fluctuating asynchronous states,
or as more or less coherent collective dynamics.
The former regime has been observed
both in balanced neural networks~\cite{bal1,bal2,bal3,ullner2020} and
in purely inhibitory networks subject to an external excitatory drive~\cite{wolf,bal4}. 
Instances of collective dynamics are discussed in~\cite{bal1,brunel2000,ostojic,ullner2018,matteo,bi2020}.
In~\cite{matteo} the authors developed a mean field (MF) formulation for a sparse balanced inhibitory network
of quadratic integrate-and-fire (QIF) neurons \cite{ermentrout1986} based on the low-dimensional reduction
methodology introduced in \cite{montbrio2015}. The idea was to
map the disorder due to the randomly distributed connections onto a quenched random distribution of
the synaptic couplings, neglecting the current fluctuations present in sparse networks \cite{brunel1999}.
However, this MF approach failed to reproduce the emergence of COs observed in the direct numerical simulations
of the network, implicitly pointing to the essential role of endogenous fluctuations in sustaining of the collective behavior.
Motivated by this failure, in this article, we revisit various MF approaches to capture
the transition from asynchronous dynamics to 
COs in sparse balanced inhibitory QIF networks with homogeneous and heterogeneous degrees distributions.

In Section II, we define the network model and introduce the relevant microscopic and macroscopic 
indicators employed to characterize the dynamical evolution.
In Section III, we approximate the network dynamics in terms of a
(nonlinear) Fokker-Planck  equation (FPE), based on the assumption of a self-consistent irregular neural activity. 
The FPE is then handled into two different way.
First, after introducing a phase representation of the neuron variable, an expansion in Fourier modes 
is considered.
In heterogeneous networks, the distribution of connectivities must also be included. However, 
under the assumption of a Lorentzian distribution, this variability can be handled without increasing the computational complexity.
A second approach is also illustrated, based on the expansion of the
probability density into circular cumulants (CCs)~\cite{tyulkina2018}.
This method very effective: as shown in the following sections, a few cumulants (actually two) provide a fairly accurate representation of
the network dynamics, including the collective periodic oscillations.
Section IV is devoted to a detailed description of homogeneous networks, 
starting from the scaling analysis of the firing rate of the asynchronous regime for a vanishingly small external current.
The linear stability of the asynchronous regimes is also performed, obtaining fairly good
estimates for the onset of COs as testified by the comparison with direct numerical simulations.
Section V is focused on the emergence of collective dynamical behavior in heterogeneous sparse networks. The role of various control 
parameters is explored: input current, average connectivity, and the degree of heterogeneity.
Finally, a summary of the main achievements is reported in Section VI together with a brief discussion of the open problems.
Appendix A contains the mathematical aspects that 
render impossible to develop a self-consistent estimation for the firing rate in
heterogeneous networks.

\section{Methods}

\subsection{The network model}

We consider $N$ inhibitory pulse-coupled QIF neurons~\cite{ermentrout1986} arranged
in a random sparse balanced network. The membrane potential of each neuron
evolves according to the equations
\begin{equation}\label{eq:1}
\tau_{m} \dot{V}_{i}(t) = I + V_{i}^2(t) - \tau_m J \sum_j\epsilon_{ji} \delta(t - t_{j}^{(k)})
\end{equation}
where $\tau_{m}=15$ ms represents the membrane time constant and
$I$ is an external DC current, encompassing the effect of distal
excitatory inputs and of the internal neural excitability.
The last term is the inhibitory synaptic current,
$J$ being the synaptic coupling. The synaptic current is the linear superposition of all the instantaneous
inhibitory postsynaptic potentials (IPSPs)
$s(t)= \delta(t)$ received by the neuron $i$ from its
pre-synaptic neurons, while $t_{j}^{(k)}$ is the $k$-th spike time of the neuron $j$,
and $\epsilon_{ji}$ is the adjacency matrix of the network. In particular,
$\epsilon_{ji} = 1$  (0) if a connection from node $j$ to $i$ exists (or not) and
$k_i=\sum_j\epsilon_{ji}$ is the number of pre-synaptic neurons connected
to neuron $i$, i.e. its in-degree.

Whenever the membrane potential ${V}_{i}$ reaches infinity,
a spike is emitted and ${V}_{i}$ is reset to $-\infty$.
In absence of synaptic coupling, the QIF model displays excitable dynamics for $I <0$,
while for positive DC currents it behaves as an oscillator with period $T_0 = \pi/\sqrt{I}$.

In order to compare numerical simulations with a recent MF theory~\cite{montbrio2015,devalle2017,matteo},
we consider sparse networks where the in-degrees $k_i$ are extracted from a Lorentzian distribution
\begin{eqnarray}\label{eq:2}
L(k)= \frac{\Delta_k}{\pi[(k-K)^2+\Delta_k^2]}
\end{eqnarray}
peaked at $K$ and with a half-width at half-maximum (HWHM) $\Delta_k$.
The parameter $\Delta_k$ measures the degree of structural
heterogeneity in the network, and analogously to Erd\"os-Renyi networks
we assume the HWHM to scale as $\Delta_k=\Delta_0\sqrt{K}$. In the numerical simulations we have truncated the distribution to avoid negative in-degrees or in-degrees larger than the network size $N$.
We have verified that the probability to
go out of the boundaries, during the generation of the distribution of the in-degrees, is always small (below 3\%): 
the associated deviations affect only marginally the observed agreement between MF theory and numerical simulations.

Finally, the DC current and the synaptic coupling are assumed to scale as
\begin{equation}
I= i_0 \sqrt{K} \quad ; \quad J=g_0/\sqrt{K}
\end{equation}
as usually done in order to ensure 
a self-sustained balanced state for sufficiently large in-degrees \cite{bal1,bal2,bal3,bal4,wolf,matteo}.

The network dynamics is integrated by employing a standard Euler scheme
with an integration time step $\Delta t = \tau_m/10000$.

\subsection{Indicators}

To characterize the collective dynamics we measure the
mean membrane potential ${v}(t) = \sum_{i=1}^N V_i(t)/N = \langle V \rangle$ and the instantaneous
population firing rate $\nu(t)$, corresponding to the number
of spikes emitted per unit of time and per neuron.

In order to measure the level of coherence in the network
dynamics, a commonly used order parameter is~\cite{scholarpedia}
\begin{equation}
\rho^2 \equiv \frac{\overline{\langle V\rangle^2}-\overline{\langle V\rangle}^2}
    {\langle \overline{V^2}-\overline{V}^2\rangle} \; ;
    \label{rho}
\end{equation}
where the overbar denotes a time average, while the angular brackets denote an ensemble average.
In practice, $\rho$ is the rescaled amplitude of the standard deviation of the mean membrane potential
$v = \langle V \rangle$.
When all neurons behave in exactly the same way (perfect synchronization),
the numerator and the denominator are equal to one another and $\rho=1$. If instead, they are
independent as in an asynchronous regime, $\rho \approx 1/\sqrt{N}$ due to the central limit theorem.
In order to estimate the amplitude of collective oscillations,
we will employ also the standard deviation $\Sigma_{\nu}$ of the population firing rate $\nu(t)$.

To estimate the level of synchronization among the neurons,
we can map the membrane potentials onto phase variables, via
the standard transformation from QIF to the $\theta$-neuron model~\cite{ermentrout1986},
namely
\begin{equation}
\label{frev}
V_i = tg\left(\frac{\theta_i}{2} \right) \quad {\rm with} \quad \theta_i \in [-\pi:\pi] \enskip .
\end{equation}
The degree of synchronization can now be quantified by the
modulus $| \cdot |$ of the complex Kuramoto order parameter \cite{kura}
\begin{equation}
z_1  = \frac{1}{N} \sum_{k=1}^N {\rm e}^{i \theta_k} \quad .
\label{kura}
\end{equation}
In completely desynchronized phases $|z_1| \propto 1/\sqrt{N}$,
while partial (full) synchronization corresponds to a  finite (``1") $|z_1|$ value.

Two parameters are typically used to characterize the
microscopic activity: the average inter-spike interval (ISI) (or, equivalently the firing rate)
and the coefficient of variation $cv_i$, i.e.
the ratio between the standard deviation and the mean of the
ISIs of the spike train emitted by the $i$th neuron.
Sometimes, the average coefficient of variation, $CV = \sum_i cv_i/N$ is considered.

Time averages and fluctuations are usually estimated on time intervals
$T_s \simeq 90$ s, after discarding a transient $T_t \simeq 15$ s.

\section{Mean Field Approaches}

At a MF level, the evolution equation \eqref{eq:1} can be rewritten
for the sub-population of neurons with in-degree $k_j$ as the following
Langevin equation
\begin{equation}
\label{fre1}
\dot{V_j} = V_j^2 + A_{g_j}(t) + \sigma_{g_j}(t) \xi_j(t)
\end{equation}
where 
\begin{equation}
A_{g_j}(t) = \sqrt{K} \left[ i_0- g_j \nu(t) \right] \; ,
\label{poisson1}
\end{equation}
$\nu(t)$ being the instantaneous firing rate, while
$\{g_j\}= \{g_0 k_j/K \}$ is the effective synaptic coupling
distributed according to a Lorentzian $L(g)$
peaked at $g_0$ with HWHM $\Delta_g=\Gamma/\sqrt{K}$
and $\Gamma = \Delta_0 g_0$.
Moreover, $\xi_j$ is a $\delta$-correlated Gaussian noise with
unitary variance ($\langle \xi_j(t) \xi_m(t) \rangle = \delta_{jm} \delta(t)$).
The noise amplitude is typically estimated by assuming
that the single spike-trains are independent Poisson processes~\cite{brunel2000},
\begin{equation}
\sigma_{g_j} (t) = \sqrt{g_0 g_j \nu(t)} \; .
\label{poisson}
\end{equation}
More in general, the variance of the current fluctuations
is given by $\sigma_{g_j}^2 (t) = {\cal F} g_0 g_j \nu(t)$, where
${\cal F}$ is the Fano factor, measuring the ratio between the
variance of the spike count and its average \cite{tuckwell1988, nawrot2010}.
For a stationary renewal process, ${\cal F} = (CV)^2$, where $CV$ is the coefficient
of variation of the spike train \cite{tuckwell1988}. Therefore,
at a first level of approximation, the effect of non Poissonian distributions can be taken in account by
expressing the amplitude of the current fluctuations as
\begin{equation}
\sigma_{g_j}^{(R)} (t) = CV \sqrt{g_0 g_j \nu(t)} \quad .
\label{renewal}
\end{equation}

\subsection{Fokker-Planck Formulation}

The  Langevin equation  \eqref{fre1}  for  the  dynamics  of
the membrane potential of the sub-population with effective coupling $g$
is equivalent to a  Fokker-Planck equation  describing  the  evolution  of the probability distribution $P_g (V,t)$, 
\begin{equation}
\label{fre2}
{\partial_t P_g(V,t)} = - {\partial_V}[(V^2 + A_g(t))P_g(V,t)]
+ D_g(t){\partial^2_{V^2} P_g(V,t)}
\end{equation}
where 
\begin{equation}
\label{eq:Dg}
D_g=\frac{\sigma^2_g}{2}  = \frac{g_0 g_j \nu}{2} \; .
\end{equation}

This can be rewritten as a continuity equation,
\begin{equation}
\label{fre3}
\frac{\partial P_g(V,t)}{\partial t} = - \frac{\partial}{\partial v} F_g(V,t)
\end{equation}
where $F_g(V,t)$ represents the flux
\begin{equation}
\label{balance}
 F_g(V,t)= (V^2 + A_g)P_g(V,t) - D_g\frac{\partial P_g}{\partial V}
\end{equation}
accompanied by the boundary condition
\begin{equation}
\nu(t) = \int dg F_g (V=+\infty,t) L(g) = \int \nu_g(t) L(g)
\end{equation}
where $\nu$ is the mean firing rate, while $\nu_g(t)$ refers to the $g$-sub population.

In order to solve the FPE, we map the membrane potential onto a phase variable
via the transformation \eqref{frev}. The new PDF reads as
\begin{equation}
R_g(\theta)=P_g (V)\frac{dV}{d\theta} \qquad{\rm where} \quad \frac{dV}{d\theta}=\frac{1}{2 \cos^2{(\theta/2)}}
\label{v_theta}
\end{equation}
and the FPE~(\ref{fre2}) can be rewritten as
\begin{equation}
\label{fretheta}
{\partial_t R_g(\theta,t)} = - {\partial_\theta}\left[\psi_0(\theta)R_g(\theta,t)\right] + {\partial_\theta } \left[ Z_0(\theta){\partial_\theta R_g(\theta,t)}\right]
\end{equation}
where
\begin{eqnarray}
\psi_0(\theta) &=& (1-\cos(\theta)) +(A_g+D_g \sin(\theta))(1+\cos(\theta))  \nonumber \\
 Z_0(\theta) &=& D_g (1+\cos(\theta))^2
\end{eqnarray}
Finally,
\begin{equation}
\label{fre4}
Q_g(\theta,t) = \psi_0(\theta)R_g(\theta,t) - Z_0(\theta){\partial_\theta R_g(\theta,t)}
\end{equation}
represents the flux in the new coordinates. The flux at the threshold $\theta= \pi$ is
linked to the firing rate by the self-consistent condition
\begin{equation}
\int dg Q_g(\pi,t) L(g) = 2 \int dg R_g(\pi,t) L(g) = \nu(t) \; .
\end{equation}
Since we are now dealing with a phase variable, it is
natural to express the PDF in Fourier space,
\begin{equation}
R_g(\theta,t) = \frac{1}{2 \pi} \left[1 + \sum_{m=1}^\infty a_m(t) {\rm e}^{-i m \theta} +c.c. \right] \; .
\label{pdf_FFT}
\end{equation}
The associated Kuramoto-Daido order parameters \cite{daido1992} for the population synchronization are
given by
\begin{equation}
z_m = \int dg \ a_m \ L(g)
\label{daido}
\end{equation}
while the equations for the various modes are
\begin{eqnarray}
\dot{a}_m &=& m \left[ i  (A_g+1)  a_m + \frac{i}{2} (A_g-1) ( a_{m-1}+ a_{m+1}) \right]
\nonumber\\
-&D_g& \left[ \frac{3m^2}{2} a_m+ (m^2-\frac{m}{2}) a_{m-1} +(m^2+\frac{m}{2}) a_{m+1} \right.
\nonumber\\
  &+&\frac{m(m-1)}{4} a_{m-2} \left. +\frac{m(m+1)}{4}a_{m+2}     \right]
\label{FP_FFTT}
\end{eqnarray}
where, by definition, $a_0=1$, $a_{-m}=a_m^\ast$.

Since $g$ is distributed according to a Lorentzian law, the
heterogeneity can be exactly taken in account by averaging over the parameter $g$.
By rewriting the distribution as
\begin{equation}
L(g) = \frac{1}{2i} \left[ \frac{1}{(g-g_0) -i \Delta_g} - \frac{1}{(g-g_0)+i \Delta_g} \right] \; ,
\label{Lg}
\end{equation}
we observe that it has two complex poles at $g = g_0 \pm i \Delta_g$.
Therefore, by invoking the Cauchy's residue theorem one can estimate
explicitly the Kuramoto-Daido order parameters as
\begin{equation}
z_m = \int dg \enskip a_m(g) \enskip L(g) =  a_m(g_0-i \Delta_g)
\label{daido2}
\end{equation}
and by averaging Eq.~\eqref{FP_FFTT} over the $g$-distribution,
one can find also the dynamical equations ruling the evolution of these quantities,
\begin{widetext}
\begin{eqnarray}
\dot{z}_m &=& m \left[ (i A_{g_0}+i-\nu \Gamma ) z_m +\frac{1}{2}(i A_{g_0} -i - \nu \Gamma)(z_{m-1}+z_{m+1}) \right]
\nonumber\\
&-& D_{g_0} (1-\frac{i\Delta_g}{g_0}) \left[\frac{3 m^2}{2} z_m +(m^2 - \frac{m}{2})  z_{m-1} + (m^2 + \frac{m}{2}) z_{m+1}
+ \frac{m(m-1)}{4} z_{m-2} +\frac{m(m+1)}{4}z_{m+2} \right] \enskip .
\label{FP_ZM}
\end{eqnarray}
\end{widetext}

As shown in \cite{montbrio2015}, the population firing rate $\nu$ and the mean membrane potential $v$
can be expanded in terms of the Kuramoto-Daido order parameters, as follows:
\begin{equation}
 W \equiv \pi \nu + i v = 1 -2 \sum_{k=1}^\infty (-1)^{k+1} z_k^\ast
  \enskip .
\label{Wg}
\end{equation}

\subsection{Ott-Antonsen Ansatz}

If one neglects the fluctuations (i.e. setting $D_{g_0} = 0$), the Ott-Antonsen (OA) manifold $z_m = (z_1)^m$
is invariant and attractive~\cite{ott2008, mirollo2012}, and
Eq.~\eqref{FP_ZM} reduces to
\begin{equation}
2 \dot{z}_1=(i A_{g_0}- \nu \Gamma) [1+z_1]^2 -i [1-z_1]^2 \quad ,
\label{OA}
\end{equation}
while Eq.~\eqref{Wg} becomes the conformal transformation~\cite{montbrio2015},
\begin{equation}
z_1 = \frac{1 - W^\ast}{1+W^\ast} \,  ,
\label{CM}
\end{equation}
which relates directly the Kuramoto order parameter
with the macroscopic observables $v(t)$  and $r(t)$ describing the network dynamics.

The application of this transformation to~\eqref{OA} leads to the two following ODEs for
$v(t)$  and $\nu(t)$ \cite{matteo}:
\begin{eqnarray}
\dot{\nu} &=& \nu(2v +\Gamma/\pi) \nonumber \\
\dot{v} &=& v^2 +\sqrt{K}(i_0 -g_0 \nu) -(\pi \nu)^2 \; .
\label{montbrio}
\end{eqnarray}

These MF equations admit a unique stable solution for any parameter choice: a focus
\cite{matteo}. This contrasts with the direct numerical simulations, which instead
reveal the emergence of periodic COs for sufficiently large median in-degree $K$.
Hence, we conclude that fluctuations must be included in the MF formulation,
if we want to reproduce the macroscopic dynamics.

In spite of this intrinsic weakness, the frequency of the damped oscillations
exhibited by Eq.~\eqref{montbrio},
is very close to the frequency
$\nu_{CO}$ of the sustained COs observed in network simulations
over a wide range of parameter values ~\cite{matteo}.

\subsection{Circular Cumulants Approximation}

In the presence of {\it weak} noise, one can go beyond the OA Ansatz,
expanding the PDF into the so-called circular cumulants (CCs)~\cite{tyulkina2018}.

In Ref.~\cite{tyulkina2018} it was noticed that the Kuramoto-Daido order parameters
\begin{equation}
z_m = \int dg \int d \theta R_g(\theta,t)L(g) {\rm e}^{i m \theta} =\langle {\rm e}^{i m \theta} \rangle
\end{equation}
are the moments of the observable ${\rm e}^{i \theta}$, which can be determined
via the following moment-generating function:
\begin{equation}
F(k) = \langle \exp{(k {\rm e}^{i \theta})} \rangle \equiv \sum_{m=0}^\infty z_m \frac{k^m}{m !}
\enskip .
\label{mon_gen}
\end{equation}

Given $F(k)$, one can obtain the CCs ${\kappa}_m$ from the cumulant-generating function \cite{tyulkina2018} :
\begin{equation}
\Psi(k) = k \partial_k \ln{F(k)} \equiv \sum_{m=0}^\infty {\kappa}_m k^m
\label{circ_gen}
\enskip .
\end{equation}
By combining Eq.~\eqref{mon_gen} and \eqref{circ_gen} one can relate $z_m$ with $\kappa_m$,
\begin{equation}
\kappa_m=\frac{z_m}{(m-1)!}-\sum_{n=1}^{m-1} \frac{\kappa_n z_{m-n}}{(m-n)!} \; .
\end{equation}
Notice that the CCs $\kappa_m$ are scaled differently from the
conventional cumulants,
which would yield $\kappa_m^\prime = (m-1) ! \kappa_m$~\cite{tyulkina2018}.
The first two CCs are therefore given by
\begin{equation}
\kappa_1 = z_1 \qquad \kappa_2 = z_2 - z_1^2
\enskip .
\end{equation}

Whenever the OA Ansatz holds, i.e. when the manifold $z_m = z_1^m$ is attractive,
the generating functions can be simply expressed as:
$$F(k) = {\rm e}^{k z_1} \qquad;\qquad \Psi(k) = k z_1 \enskip ;$$
where $\kappa_1 = z_1$ is the only non zero CC.

In general, when the OA manifold is not attractive,
all CCs are non zero. However, in~\cite{tyulkina2018} it was found
that their amplitude decreases exponentially with their order, $\kappa_m \propto D_{g_0}^{m-1}$,
where $D_{g_0}$ is the noise intensity.
Therefore it makes sense to restrict the expansion to the first two CCs, in the
weak-noise limit.
Under this approximation, the Kuramoto-Daido order parameters are simply given by
\begin{equation}
z_m = z_1^m + \kappa_2 z_1^{m-2} \frac{m(m-1)}{2} \; .
\end{equation}
The second addendum on the r.h.s. can be interpreted as a correction to the OA manifold due to the noise.

The 2CCs approximation for the FPE \eqref{FP_ZM} (correct up to order ${o}(D_{g_0})$) reads as
\begin{widetext}
\begin{eqnarray}
\dot{z}_1 &=& z_1 (i A_{g_0} +i -\Gamma \nu)+H(1+\kappa_2+z_1^2)-\frac{D_{g_0}}{2} (1-i\frac{\Delta_g}{g_0})(1+z_1)^3 \\
\dot{\kappa}_2 &=& 2(i A_{g_0} + i- \Gamma \nu)\kappa_2+4 H z_1 \kappa_2- D_{g_0}(1-i\frac{\Delta_g}{g_0})
\left(\frac{1}{2} (1+z_1)^4+ 6 (1+z_1)^2 \kappa_2 \right)
\label{2CC}
 \end{eqnarray}
\end{widetext}
where
$$
H=\frac{1}{2}\left[i(A_{g_0}-1)-\Gamma \nu \right] \quad .
$$
The firing rate $\nu$ and the mean membrane potential $v$ can
be obtained from Eq.~\eqref{Wg} by restricting the sum to the first two CCs,
\begin{equation}
W^\ast =
\pi \nu-i v=\frac{1-z_1}{1+z_1} +\frac{2\kappa_2}{(1+z_1)^3}
\label{conformal_ext}
\end{equation}
this is a generalization of the conformal transformation \eqref{CM}
to a situation where the OA Ansatz is no longer valid.

\subsection{Annealed network}

In the homogeneous case ($\Delta_0=0$) we performed direct numerical simulations of the Langevin equations (\ref{fre1})-(\ref{poisson1}). We considered $N_{ann}$ uncoupled neurons whose membrane potential follows Eq.(\ref{fre1}), integrating their dynamics with an Euler integration scheme with integration step $dt=5\cdot10^{-5}$. The population firing rate $\nu(t)$ at time $t$ is estimated self-consistently by counting the spikes emitted by the $N_{ann}$ neurons in the preceding time interval of duration $\Delta_t=0.01$. The advantage of these simulations is that we can reach arbitrarily large values of the connectivity $K$. This integration scheme has been employed to study the dependence of COs' features on the median in-degree $K$, as shown in Fig.~\ref{fscal}.   

\section{Homogeneous Case}

We now restrict our analysis to the homogeneous case, where the in-degree is equal to $K$ for all the neurons,
i.e. $g_j = g_0 \enskip \forall j$ and $\Delta_g = \Delta_0 = 0$. Fluctuations are still expected
since each neuron receives
inputs from a different randomly chosen set of $K$ pre-synaptic neurons \cite{brunel2000}.
In this case the Kuramoto-Daido order parameters coincide with the coefficients
of the Fourier expansion of the PDF, i.e.  $z_m \equiv a_m (g_0)$.

\subsection{Asynchronous State}

\subsubsection{Fourier space representation of the FPE}

The asynchronous state is identified by a stationary PDF, which can be obtained by solving
the FPE~\eqref{FP_FFTT} in Fourier space, truncating the hierarchy at some order $M$.

\begin{figure}
\centerline{\includegraphics[width=0.45\textwidth]{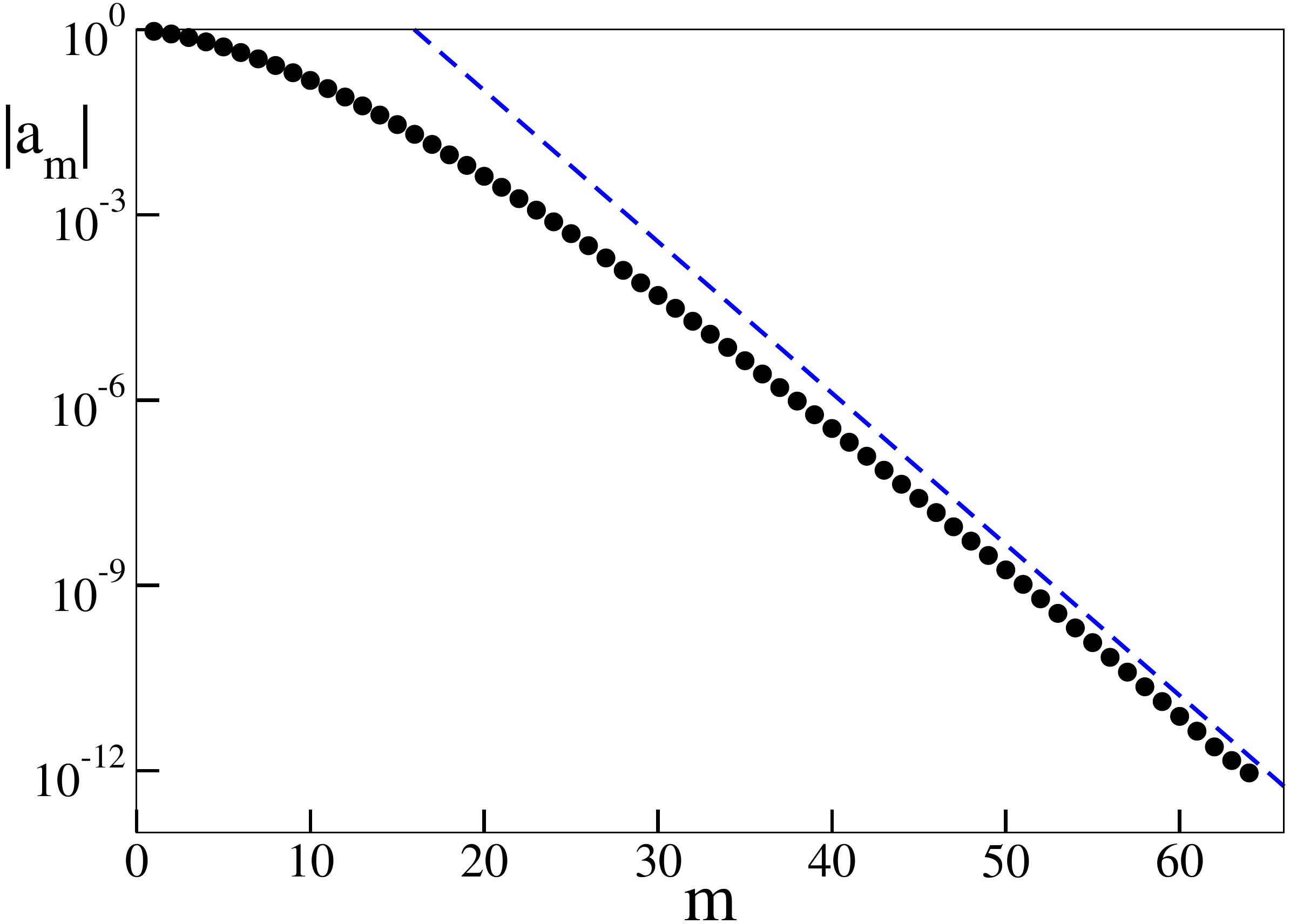}}
\caption{Modulus of the Fourier coefficients $a_m$. The blue dashed
curve corresponds to an the exponential decay law with exponent $-0.564$.
Parameters: $i_0 = 0.006$, $g_0=1$, $\Delta_g=0$ and $K=40$.
}
\label{f1}
\end{figure}

In practice we have solved iteratively the
linear (in the coefficients $a_m$ and $a^\ast_m$) system \eqref{FP_FFTT}, accompanied by the
nonlinear consistency condition
\begin{equation}
\nu^{(0)} =  2 R^{(0)}(\pi) = \frac{ 1 + \sum_{m=1}^\infty (-1)^{m+1}(a_m^{(0)} + {a^{(0)}_m}^\ast )}{\pi}
\label{nu0}
\end{equation}
where the superscript $(0)$ means that we refer to the stationary state.
The Fourier spectrum of coefficients is shown in Fig.~\ref{f1} for a certain choice of parameter values;
the amplitude $|a_m|$ decays exponentially with $m$
with an exponent approaching $\simeq -0.564$ for sufficiently large $m$. Hence,
the truncation to $M=64$ Fourier modes is very accurate, since it amounts
to neglecting terms ${\cal O}(10^{-12})$, and indeed we do not observe any appreciable difference by increasing $M$.

\begin{figure}
\centerline{\includegraphics[width=0.45\textwidth]{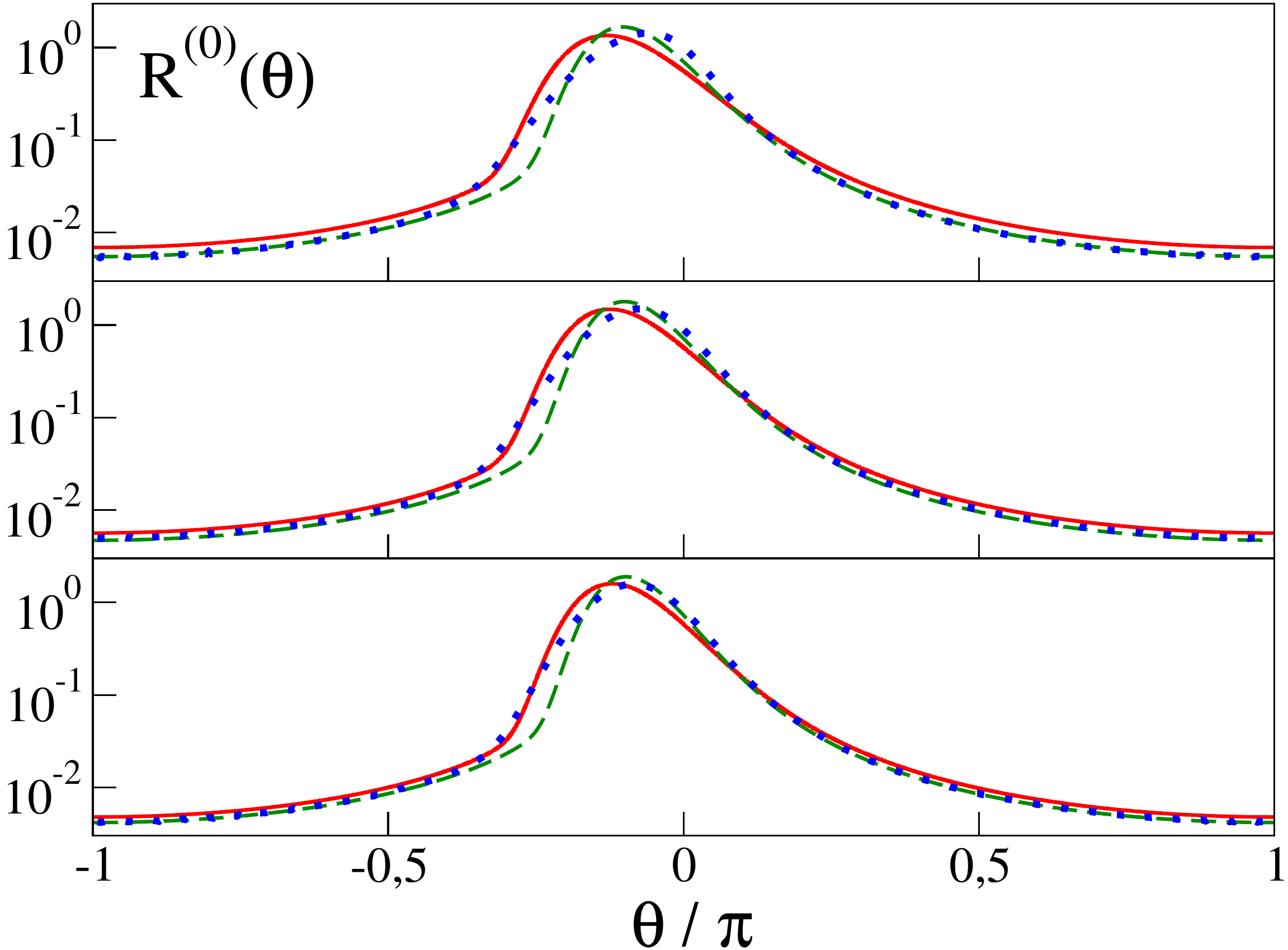}}
\caption{Stationary PDFs $R^{(0)}$ versus the angle $\theta$ estimated numerically from the
network simulations (blue dotted line) and theoretically from the FPE truncated at $M=64$
by estimating the current fluctuations
within the Poisson approximation Eq. \eqref{poisson} (red solid line) and
within the renewal approximation Eq. \eqref{renewal} with $CV=0.8$ (green dashed line).
From top to bottom $K=20$, 40
and 80. The numerical data are obtained for a network of size $N= 16000$ and by averaging over $20$ different neurons.
Parameters: $i_0 = 0.006$, $g_0=1$ and $\Delta_g=0$.
}
\label{f2}
\end{figure}

In Fig.~\ref{f2} we display the stationary PDF for three different in-degrees ($K=20$, 40, and 80).
The blue dotted lines have been obtained by simulating a network of $N=16000$ neurons.
We have verified that finite-size corrections are negligible.
Moreover, we found that an average over 20 neurons suffices to reproduce the PDF  of the whole
ensemble.
The red solid and green dashed curves have been obtained by solving the FPE equation under the Poisson (Eq.~\eqref{poisson}),
resp. renewal (Eq.~\ref{renewal}) approximation for the synaptic-current fluctuations $\sigma_g$.

The two MF theoretical curves reproduce fairly well the numerical results.
The main differences concern the peak of $R^{(0)}(\theta)$: the theoretical distributions are slightly shifted to the left,
although the shift reduces upon increasing $K$, as expected for a MF theory.
On the other hand, the PDF tail is captured quite well, and so is the
average firing rate $\nu^{(0)}= 2 R^{(0)}(\pi)$, as reported in Table~\ref{t1}.
The renewal approximation (under the assumption of $CV=0.8$ as observed in the numerical simulations) reveals a better
agreement with the direct simulations.

\begin{table}
 \begin{tabular}{||c || c |  c c | c||}
 \hline
$K$ & $\overline{\langle \nu \rangle}$ & $\nu^{(0)}$ & $\nu^{(0)}$ & $\nu_{g_0}$ \\
\hline
&  &  FPE(P) & FPE(R) & 2 CCs \\ [0.5ex]
 \hline\hline
 20 & 0.0114 &  0.0138 & 0.0110 & 0.0129 \\
 \hline
 40 & 0.0100 &   0.0112 & 0.0094 & 0.0105 \\
 \hline
 80 & 0.0089 &  0.0096 &  0.0084 & 0.0089\\
 \hline
\end{tabular}
\caption{Average population firing rate versus the in-degree $K$ for asynchronous dynamics. 
The second column reports $\overline{\langle \nu \rangle}$
as estimated by averaging the activity of a network of $N=16000$ neurons;
the average is performed over all neurons and in time.
The third and fourth columns report the firing rate obtained from the
stationary PDF, namely $\nu^{(0)} = 2 R^{(0)}(\pi)$.
More precisely, the third (fourth) row
displays the MF results obtained from the self-consistent solution of the
stationary Eq.~\eqref{FP_FFTT} for $M=64$ under the 
Poisson approximation (within the renewal approximation with $CV=0.8$ ).
The fifth columns refers to the 2CCs approximation:
the population firing rate $\nu_{g_0}$ has been estimated using
the expression~\eqref{conformal_ext}.
Parameters $i_0 = 0.006$, $g_0=1$ and $\Delta_g=0$.
}
\label{t1}
\end{table}

\subsubsection{Expansion of the FPE in CCs}

\begin{figure*}
\includegraphics[width=0.45\textwidth]{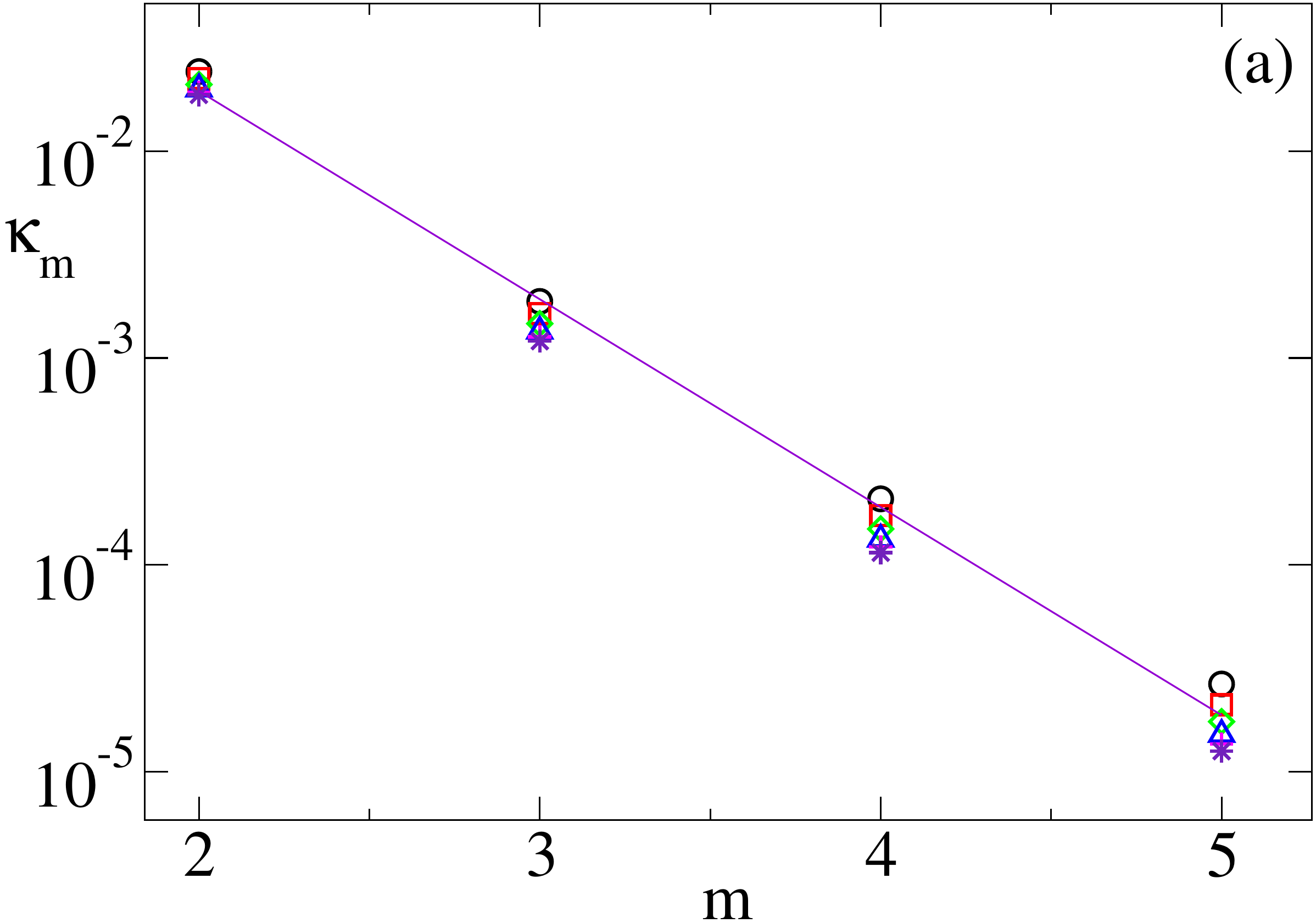}
\includegraphics[width=0.45\textwidth]{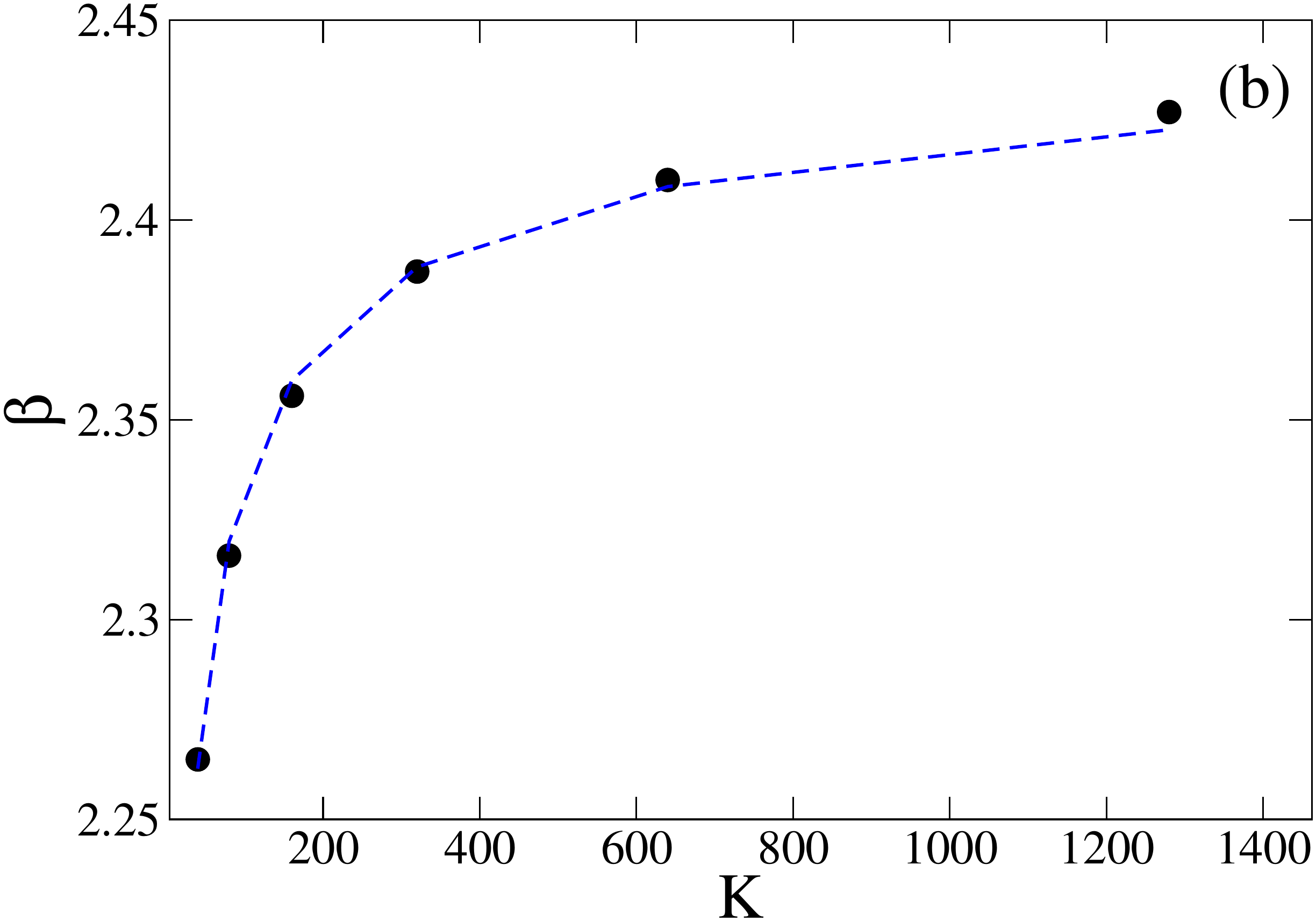}
\caption{(a) Circular cumulants $\kappa_m$ as a function of their degree $m$
for different values of the median in-degree $K$: $K=40$ (black circles),
$K=80$ (red squares), $K=160$ (green diamonds), $K=320$ (blue pluses),
$K=640$ (magenta crosses), $K=1280$ (violet asterisks). The violet solid line refers 
to an exponential fitting $\kappa_m \simeq \mathrm{e}^{-\beta m}$ of the data for $K=80$,
the exponent $\beta = 2.3$ in this case.
(b) Dependence of the exponent $\beta$ versus $K$. The blue dashed line refers to a fitting
$\beta \approx A_0 \left[ 1 - \frac{1}{2 \sqrt{K}} \right]$ where $A_0 = 2.457$. The cumulants are 
estimated from the stationary solution $a_m^{(0)}$ of the FPE Eq. \eqref{FP_FFTT} with $M=64$.
Parameters: $i_0 = 0.006$, $g_0=1$ and $\Delta_g=0$.
}
\label{f3}
\end{figure*}

We now consider the expansion in CCs.
As expected, $\kappa_1 \simeq {\cal O} (1)$, while the higher-order
CCs decrease exponentially, $\kappa_m \simeq \mathrm{e}^{-\beta(K) m}$
(see Fig.~\ref{f3}(a), where $\kappa_m$ is plotted for a few different
connectivities). The dependence on $K$ is rather weak and can be
appreciated in panel (b), where $\beta$ is plotted versus $K$.
The dependence is fitted very well by the empirical law
$\beta \approx A_0 \left[ 1 - \frac{1}{2 \sqrt{K}} \right]$
with $A_0 = 2.457$.
This result implies that $\kappa_m$ stays finite for any $m$ in the limit
$K\to\infty$, thus confirming that the self-generated noise is still relevant in perfectly balanced states. 

In~\cite{tyulkina2018}, it was found that $\beta = -\ln D_{g_0}$. Here,
estimating $D_{g_0}$ from Eq.~\eqref{eq:Dg}, with $\nu^{(0)}$ obtained
from the stationary solution of the FPE, we find the slower dependence
$\beta \approx 0.598 - 0.321 \ln D_{g_0}$.

Altogether, the fast decrease of the higher-order CCs suggest that the first two cumulants
should suffice to reproduce the observed phenomenology.
A first evidence of the validity of the 2CCs approximation comes from the
average firing rate $\nu$ obtained by employing Eq.~\eqref{conformal_ext}. The data
reported in Table~\ref{t1} indeed show that the 2CC approximation (last column)
are in good agreement with the direct numerical simulations.

\subsubsection{Self-consistent solution for the average firing rate}

Here we return to the original formulation of the FPE (see Eq.~\eqref{balance}).
The firing rate of a sub-population with effective coupling $g_0$ can be determined by solving the
linear differential equation
\[
\nu_{g_0} =(A_{g_0}+V^2)P_{g_0}-D_{g_0}\frac{\partial P_{g_0}}{\partial V} \quad .
\]
The stationary PDF $P_{g_0}(V)$ can be derived by employing the method of variation of the constants,
namely
\[
P_{g_0}(V)=\frac{\nu_{g_0}}{D_{g_0}}\int\limits_V^{+\infty}\mathrm{d}U\,e^{-\frac{A_{g_0}}{D_{g_0}}(U-V)-\frac{U^3-V^3}{3D_{g_0}}} \quad
\]
Hence, the firing rate $\nu_{g_0}$ can be obtained by normalizing the PDF $P_{g_0}(V)$ (see Refs.~\cite{brunel2003,lindner2003}),
\begin{equation}
1 = \int\limits_{-\infty}^{+\infty}\mathrm{d}V\,P_{g_0}(V) =
\frac{\nu_{g_0}\sqrt{\pi}}{\sqrt{D_{g_0}}} \int\limits_{0}^{+\infty}\frac{\mathrm{d}y}{\sqrt{y}}  
e^{\frac{- A_{g_0} y-\frac{y^3}{12}}{D_{g_0}}}
\enskip ;
\label{nug_imp}
\end{equation}

The integrals appearing in \eqref{nug_imp} can be analytically estimated, leading to 
\begin{widetext}
\begin{equation}
\nu_{g_0}=D_{g_0}^{1/3}\mathcal{R}(\xi)=
\left\{
\begin{array}{cc}
\displaystyle
\frac{-9D_{g_0}/(4\pi^2A_{g_0})}
{ \mathrm{I}^2_{\frac13}(\chi_-) + \mathrm{I}^2_{-\frac13}(\chi_-)+
  \mathrm{I}_{\frac13}(\chi_-)\,\mathrm{I}_{-\frac13}(\chi_-)}, & A_{g_0} < 0 \enskip ; \\[30pt]
\displaystyle
\frac{9D_{g_0}/(4\pi^2A_{g_0})}
{ \mathrm{J}^2_{\frac13}(\chi_+) + \mathrm{J}^2_{-\frac13}(\chi_+)-
  \mathrm{J}_{\frac13}(\chi_+)\,\mathrm{J}_{-\frac13}(\chi_+)}, & A_{g_0} > 0 \enskip ;
\end{array}
\right.
\label{eq:103}
\end{equation}
\end{widetext}
where $\xi = A_{g_0}/D_{g_0}^{2/3}$,  $\chi_{\pm} = 2(\pm\xi)^{3/2}/3$,
while $\mathrm{J}_n$ and $\mathrm{I}_n$ are the $n$-th
order Bessel function of the first kind and the modified one, respectively.
In practice, the two expressions for positive and negative $A_{g_0}$ are the analytic
continuation of one another, but we prefer to keep an explicit formulation with real values
in order to avoid incorrect choices of these two-sheet analytic functions \cite{cont}.

Now, by recalling that $D_{g_0} = g_0^2 \nu_{g_0}/2$, one can turn Eq.~\eqref{eq:103} into an equation linking
$\nu_g$ with $\xi$
\begin{equation}
\nu_{g_0}(\xi)=\frac{g_0}{\sqrt{2}}[\mathcal{R}(\xi)]^{3/2} \; .
\label{eq:nuxi}
\end{equation}
Self-consistency is finally imposed by rewriting Eq.~\eqref{poisson1} as
\begin{equation}
\xi = \frac{\sqrt{K}}{\nu_{g_0}^{2/3}(\xi)} \left[ i_0- g_0 \nu_{g_0}(\xi) \right] \; .
\label{eq:201}
\end{equation}
This equation allows determining the unknown $\xi$, which, in turn, allows finding the firing rate
from \eqref{eq:nuxi}.

\begin{figure*}
\includegraphics[width=0.32\textwidth,clip]{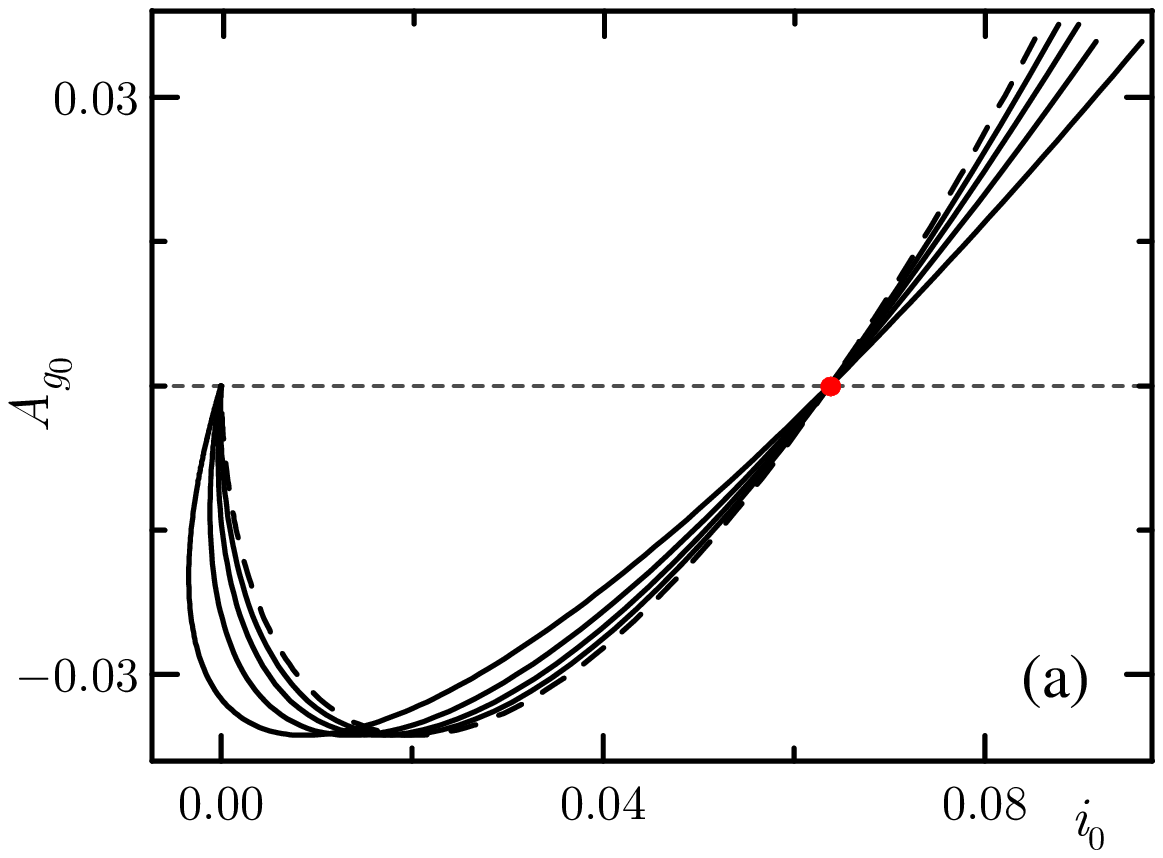}
\includegraphics[width=0.32\textwidth]{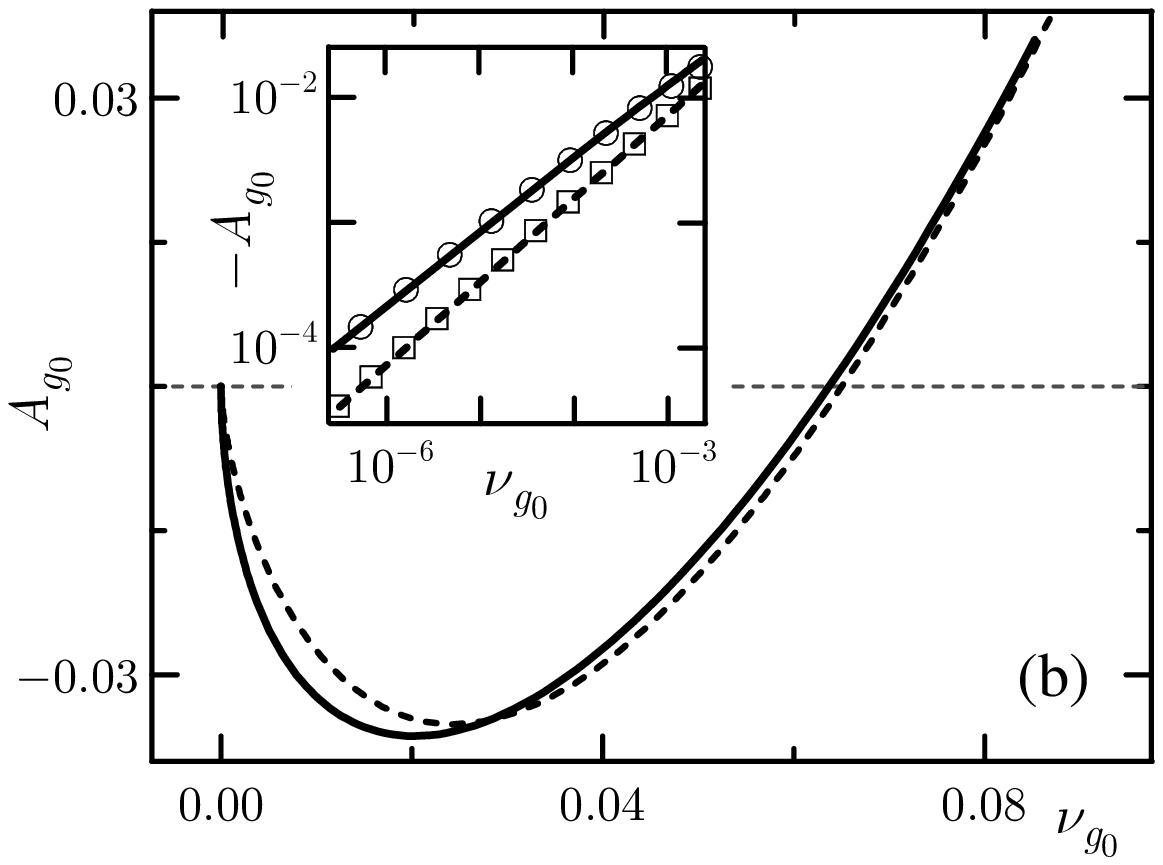}
	\includegraphics[width=0.32\textwidth]{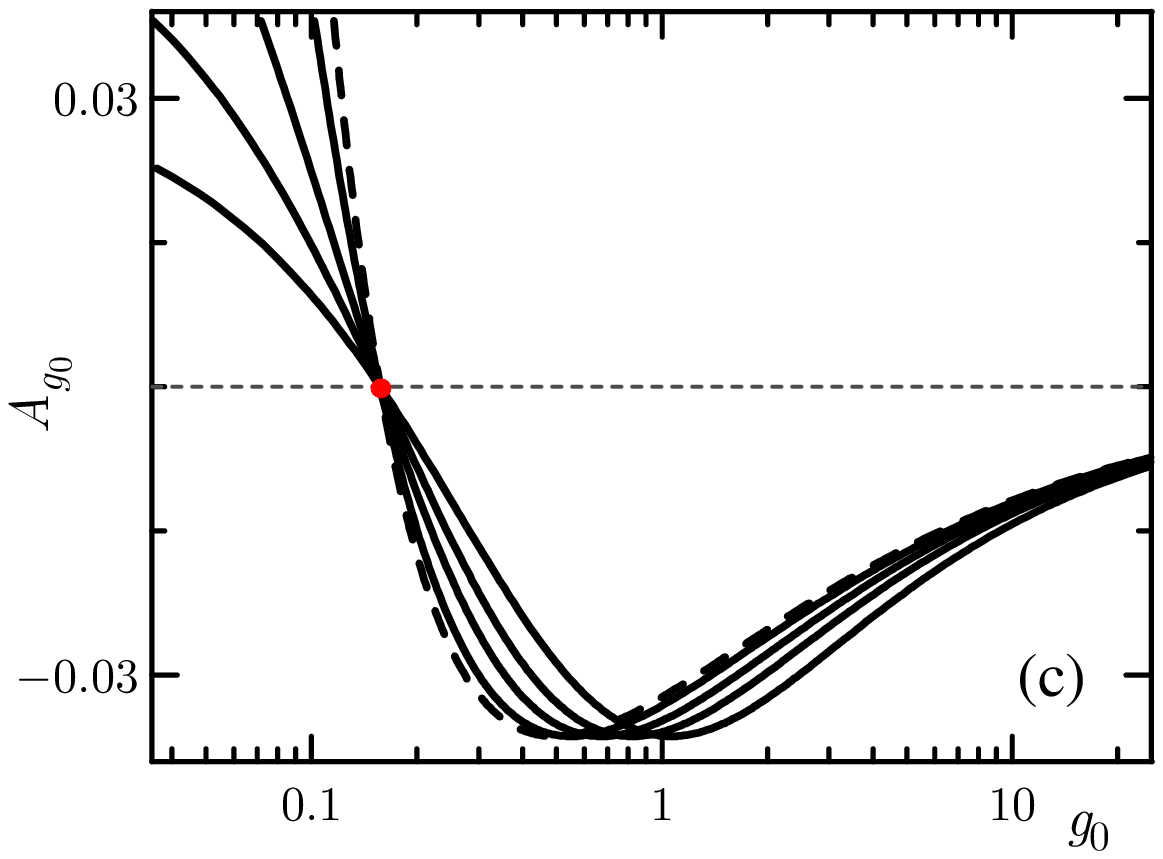}
\caption{(a) $A_{g_0}$ versus $i_0$ for a homogeneous network ($\Delta_0=0$) for $K=10$, $30$, $100$, $1000$ (solid lines from left to right on the bottom left of the panel) and for $K=\infty$ (dashed line). The solution is calculated in parametric form~\eqref{eq:201}.
The red circle corresponds to the balanced solution $i_{*}$ (Eq.~\eqref{eq:202}). 
(b) $A_{g_0}$ versus $\nu_{g_0}$ for $K=\infty$ : solid line is the exact result,
the dashed one corresponds to the 2CCs approximation. In the inset the exact solution is 
shown as circles and the 2CCs approximation as squares, 
while the solid line refers to a fitting of the exact data with $-A_{g_0} \propto \nu_{g_0}^{2/3}\ln\nu_{g_0}$ 
and the dashed one to a fitting of the 2CCs solution with 
$-A_{g_0}\propto\nu_{g_0}^{2/3}$. 
(c)  $A_{g_0}$ versus ${g_0}$; the dashed line refers to $K=\infty$, while
the solid ones to $K=10$, $30$, $100$, $1000$ (from left to right on the upper left of the panel). 
In panel (a) and (b)  $g_0=1$, while in panel (c) $i_0=0.01$. 
}
\label{fig2}
\end{figure*}

In Fig.~\ref{fig2}(a), we report $A_{g_0}$ versus the input current $i_0$.
$A_{g_0}/\sqrt{K}$ represents the unbalance: the 
deviation of the firing rate $\nu_{g_0}$ from the balanced regime $i_0/g_0$.
Upon increasing $K$, 
$A_{g_0}(i_0)$ stays finite and converges to a limiting shape (see the different curves).
Thus, we can conclude that when $K\to \infty$ perfect balance is eventually attained.
Interestingly, there exists a special current $i_*$, whose
corresponding state is perfectly balanced for any $K$ value.
Its value can be identified from
Eq.~\eqref{eq:201} by imposing the condition $\xi=0$,
\begin{equation}
i_* = \frac{g_0^2}{\sqrt{2}}[ \mathcal{R}(0)]^{3/2} =
\frac{9g_0^2}{\sqrt{2}}\left(\frac{\Gamma(2/3)}{2\pi}\right)^3 = 0.0637...\cdot g_0^2 \,
\label{eq:202}
\end{equation}
(we have made use of the definition of the Bessel functions).
For $i_0 > i_*$, the asynchronous state becomes increasingly mean-driven indicating that the inhibitory feedback due
to the coupling is less able to counterbalance the excitatory external current.
For $i_0 < i_*$, the dynamics is instead fluctuation-driven. The negative unbalance is maximal for $i_0\approx 0.02$.
This is consistent since the response of the network (in whatever direction) is expected to decrease with $i_0$.

Much less obvious is that for low $K$, asynchronous states seem to exist for negative currents.
However, as argued in the conclusions, there are strong reasons to disbelieve that the white noise assumption
is valid in such a circumstance. Hence, we do not further comment on this feature.

In Fig.~\ref{fig2}(c), we explore the dependence of $A_{g_0}$ on the synaptic coupling for $i_0=0.01$.
We see that the dynamics is fluctuation- (mean-) driven for large (small) $g_0$.
This is not only reasonable (since the coupling is inhibitory), but also agrees with the 
results reported in \cite{lerchner2006}.

\begin{figure}[!b]
\includegraphics[width=0.375\textwidth]{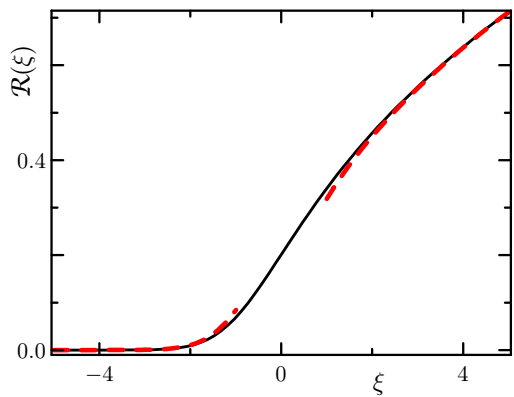}
\caption{
The function $R(\xi)$ versus $\xi$ for the $g$-sub-population of QIF neurons.
Black solid line: the exact expression ~(\ref{eq:103}); red dashed lines: the asymptotic approximation~(\ref{eq:104b}).
}
  \label{fig1}
\end{figure}

Finally, we derive some approximate analytic expressions, useful both to establish the
scaling behavior for small currents and to 
compare with the CC approximation discussed in the following section.
Let us start plotting $\mathcal{R}$ versus $\xi$ in Fig.~\ref{fig1} (see the solid black curve):
it vanishes for $\xi \to -\infty$, while it diverges for $\xi \to \infty$.
By recalling that $\nu \simeq \mathcal{R}^{2/3}$ (see Eq.~\eqref{eq:nuxi}) the same conclusion
holds for $\nu$, meaning that a small firing rate corresponds to a very negative $\xi$, while $\nu\gg 1$ corresponds
to large and positive $\xi$.
In these two limits, the following asymptotic formulas hold
\begin{equation}
\mathcal{R}(\xi)\approx
\left\{
\begin{array}{cc}
\frac{\sqrt{-\xi}}{\pi}\exp\left(-\frac{4(-\xi)^{3/2}}{3}\right), & \xi \ll -1
\\[15pt]
\frac{\sqrt{\xi}}{\pi}\,,& \xi \gg 1 \enskip.
\end{array}
\right.
\label{eq:104b}
\end{equation}
The upper expression is basically the Kramers escape rate for the overdamped dynamics of a particle in a potential well of height
$\propto (-A_{g_0})^{3/2}$ \cite{lindner2003}. The lower expression refers to the activity of an isolated supra-threshold QIF neuron.
The validity of the two expressions can be appreciated in Fig.~\ref{fig1} (see the two red dashed curves).

For $\xi \ll -1$ (i.e for small current $i_0$), Eq.~\eqref{eq:nuxi} implies
\begin{equation}
\nu_{g_0}\approx\frac{g_0 (-\xi)^{3/4}}{2^{1/2}\pi^{3/2}}\exp[-2(-\xi)^{3/2}]\,.
\label{eq:204}
\end{equation}

Upon taking the logarithm of both sides and neglecting the prefactor of the exponential term,
\begin{equation}
\ln \nu_{g_0} \simeq -2 (-\xi)^{3/2}  \; .
\label{eq:nuxi2}
\end{equation}
This equation allows eliminating $\xi$ from Eq.~\eqref{eq:201}, obtaining
\begin{equation}
i_0 = g_0 \nu_{g_0} - \frac{\nu_{g_0}^{2/3}}{\sqrt{K}} \left [\frac{\ln (1/\nu_{g_0})}{2} \right ]^{2/3}
\label{eq:205}
\end{equation}
This equation is valid in the limit of a small $i_0$,
as it is clearly appreciable from the inset of  Fig.~\ref{fig2}(b).

\subsubsection{2CCs expansion}

The FPE is a functional equation. It is worth exploring whether the CC expansion is able to reproduce its main properties.
In this section, we test the correctness of the 2CC approximation, with reference to the stationary solution.
The asynchronous state $z_1^{(0)}$, $k_2^{(0)}$ is obtained by looking for the stationary solution
of Eqs.~\eqref{2CC},
where $A_{g_0} = \sqrt{K} \left[ i_0- g_0 \nu^{(0)} \right]$ and $D_{g_0} =g_0^2 \nu^{(0)}/2$,
while the firing rate is determined by imposing the self-consistent condition (see Eq.~\eqref{conformal_ext})
\begin{equation}
 \nu^{(0)}= \frac {1}{\pi}
 Re \left\{ \left[ \frac{1-z^{(0)}_1}{1+z^{(0)}_1} +\frac{2\kappa^{(0)}_2}{(1+z^{(0)}_1)^3} \right] \right\}
 \quad .
\label{r0}
\end{equation}
The quality of the 2CCs approximation can be appreciated from  Fig.~\ref{fig2}(b) where
we report $A_{g_0}$ versus $\nu_{g_0}$ and we compare the exact solution (solid line) with the 2CCs approximation (dashed line).
The approximation captures reasonably well the behavior of $A_{g_0}$ over the whole range and is
particularly accurate for large $\nu_{g_0}$. In the inset, $|A_{g_0}|$ is reported for small $\nu_{g_0}$;
in this range, the exact solution scales as  $|A_{g_0}| \simeq \nu_{g_0}^{2/3} \ln \nu_{g_0}$, as previously noticed,
while the 2CCs approximation gives a scaling $|A_{g_0}| \simeq \nu_{g_0}^{2/3}$, without logarithmic corrections. 

\begin{figure}[!]
\includegraphics[width=0.4\textwidth]{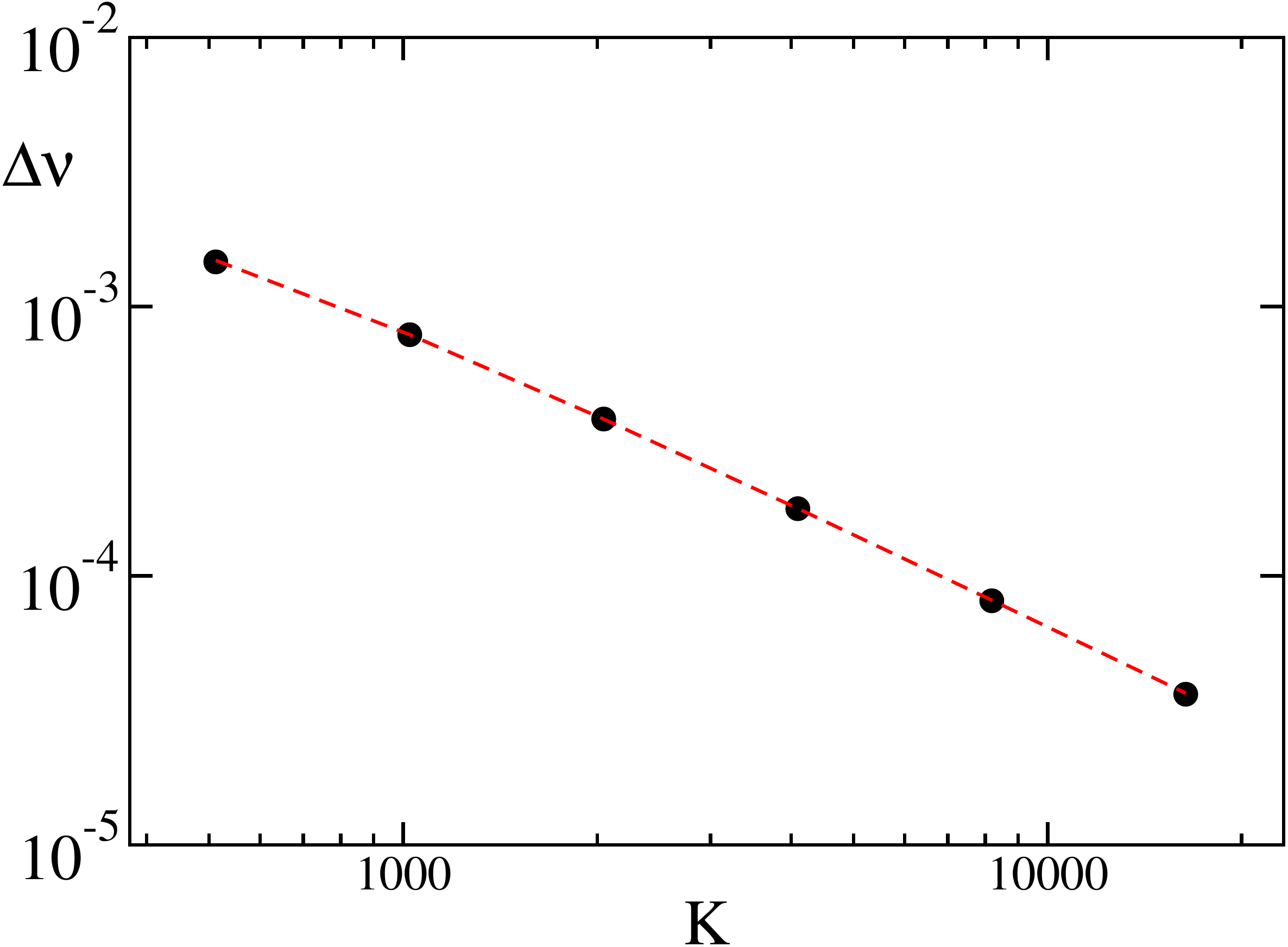}
\caption{Deviation $\Delta \nu$ from perfect balance versus the connectivity $K$ 
estimated from the stationary solutions of Eqs.~\eqref{2CC}
within the 2CCs approximation, under the assumption that the external current is $I= 7.7/\sqrt{K}$
(filled black circles).  The red dashed line corresponds to a fitting to the data
of the type $\Delta \nu = 3.08 K^{-7/6} + 168 K^2$.
 }
\label{fig:dev}
\end{figure}

Finally, we have investigated the weak-current limit, assuming
\begin{equation}
i_0 = \frac{i_w}{K}
\end{equation}
equivalent to $I = i_w/\sqrt{K}$.
The scaling analysis carried out in the end of the previous section implies that the
last term in Eq.~\eqref{eq:205} is negligible and accordingly that
$\nu \simeq i_0 \simeq \mathcal{O}(1/K)$. Moreover, it is easily seen that
\begin{equation}
\Delta\nu = \nu_g g_0 - i_0 \simeq \frac{(\ln K)^{2/3}}{K^{7/6}}
\label{dnu}
\end{equation}
Separately, we have determined the firing rate predicted by the 2CCs model, from 
the stationary solution of Eqs.~\eqref{2CC}.
The results for $i_w=7.7$ are presented in Fig.~\ref{fig:dev}, where we see that $\Delta \nu$ scales
very accurately as $K^{-7/6}$, the higher order corrections being of $1/K^2$ type.
Once again we can conclude that the 2CCs model is able to capture the leading behavior, but fails to reproduce
the logarithmic correction.  Not too bad for such a simple model.

\subsection{Linear Stability of the Asynchronous State}

\subsubsection{Fokker-Planck Formulation}

The stability of the asynchronous state can be assessed by linearizing Eq.~\eqref{FP_FFTT}
around the stationary solution $\{a_m^{(0)}\}$,
\begin{widetext}
\begin{eqnarray}
\delta \dot{a}_m &=& m \left[(i A_{g_0}^{(0)}+i) \delta a_m +\frac{1}{2}(i A_{g_0}^{(0)}-i)(\delta a_{m-1}+\delta a_{m+1}) \right]
+ i m \delta A_{g_0} \left[a^{(0)}_m +\frac{a^{(0)}_{m-1}+a^{(0)}_{m+1})}{2} \right]
\nonumber\\
&-& D_{g_0}^{(0)}\left[\frac{3 m^2}{2} \delta a_m +(m^2 - \frac{m}{2}) \delta a_{m-1} + (m^2 + \frac{m}{2}) \delta a_{m+1}
+ \frac{m(m-1)}{4} \delta a_{m-2} +\frac{m(m+1)}{4} \delta a_{m+2} \right] \nonumber\\
&-& \delta D_{g_0} \left[\frac{3 m^2}{2} a^{(0)}_m +(m^2 - \frac{m}{2}) a^{(0)}_{m-1} + (m^2 + \frac{m}{2})
a^{(0)}_{m+1} + \frac{m(m-1)}{4} a^{(0)}_{m-2} +\frac{m(m+1)}{4} a^{(0)}_{m+2} \right] \quad;
\label{FP_LIN}
\end{eqnarray}
\end{widetext}
where $A_{g_0}^{(0)}$ and $D_{g_0}^{(0)}$ are determined by inserting the firing rate
as from Eq.~\eqref{nu0}, so that
\begin{eqnarray}
\delta  A_{g_0} &=& - \sqrt{K} g_0 \delta \nu \quad, \quad \delta D_{g_0} = \frac{g_0^2}{2} \delta \nu \\
\delta \nu &=& \frac{1}{\pi} \sum_{m=1}^\infty (-1)^{m+1} (\delta a_m + \delta a_m^\ast) \quad .
\label{lincoef}
\end{eqnarray}
The system~\eqref{FP_LIN} has been solved by employing the usual Ansatz
$\delta {\bf a} (t) = {\rm e}^{i \lambda_k t} \delta {\bf a}(0)$, and truncating the hierarchy at order $M$,
so that $\delta {\bf a} = (\delta a_1, \delta a_2, \dots, \delta a_M)$ is an $M$-dimensional vector.
Hence, the problem amounts to diagonalizing an $M\times M$ real matrix.

\begin{figure*}
\includegraphics[width=0.7\textwidth]{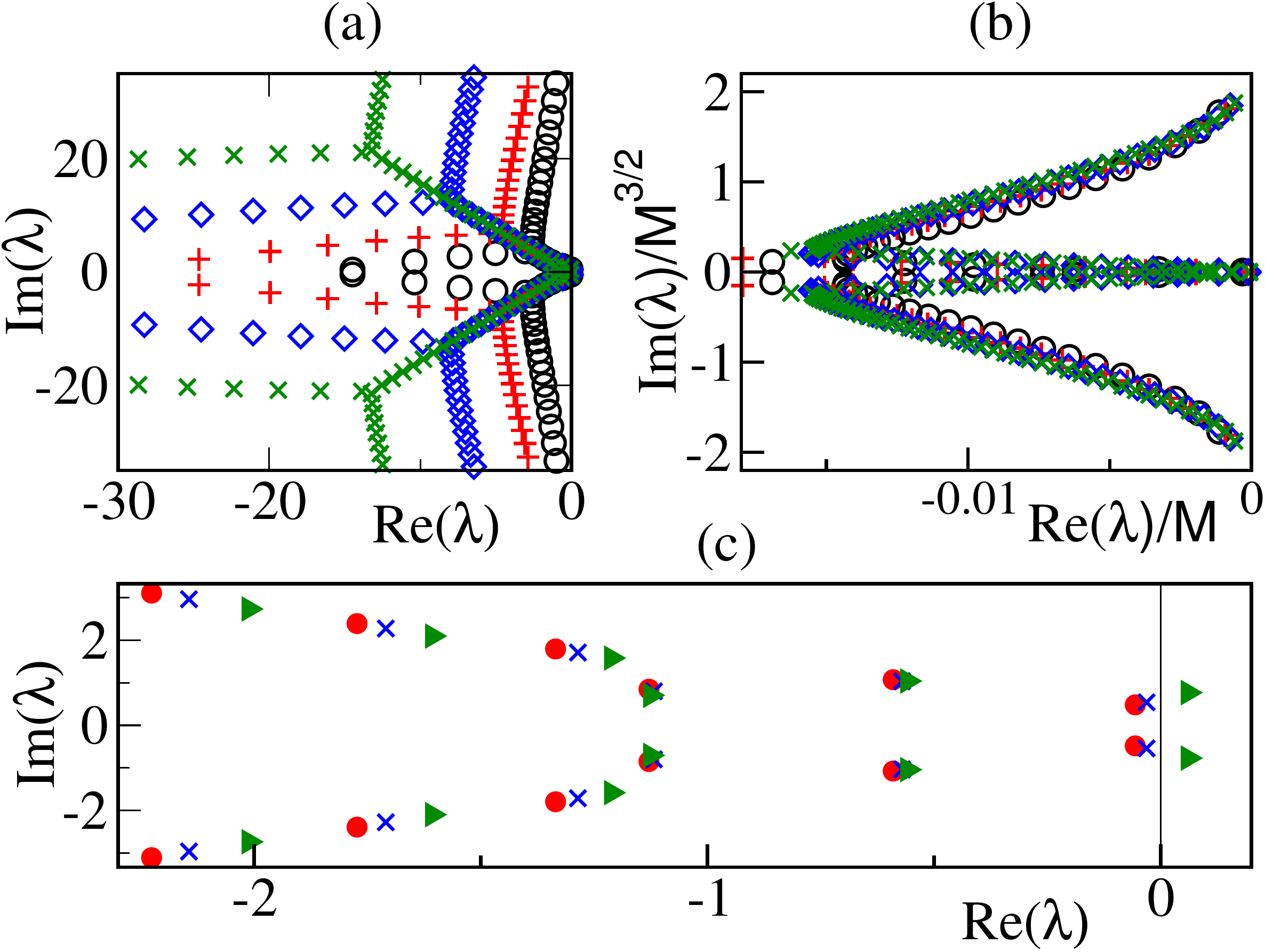}
\caption{Real and imaginary part of the eigenvalues $\{\lambda_k\}$ of the asynchronous state
for $i_0=0.006$, $g_0=1$ in a homogeneous network.
(a) Data refer to $K=40$ and different truncations of the Fokker-Planck equation
in Fourier space: namely, $M=32$ (black circles), $M=45$ (red pluses), $M=64$ (blue diamonds) and $M=90$ (green
crosses). Only a fraction of the exponents are represented.
(b) same data as in panel (a) after a suitable rescaling to show the $M$-dependence of the spurious exponents.
(c) The most relevant exponents for three different connectivities: $K=80$ (red circles), $K=160$ (blue crosses), and
$K=1600$ (green diamonds).
}
\label{f4}
\end{figure*}

The resulting spectra for a Poissonian noise are displayed in Fig.~\ref{f4}.
Each spectrum is composed of pairs of complex-conjugate eigenvalues (the matrix is real) and,
therefore, symmetric with respect to the axis $\mathrm{Im}(\lambda) = 0$.
The spectra reported in panel (a) are obtained for $K=40$, but different numbers of Fourier modes
(circles, pluses, diamonds, and crosses correspond to $M=32$, 45, 64, and 90, respectively).
There, we recognize three different branches: an almost horizontal, vertical, and a tilted one.
Only along the last one we see an overlap of the different spectra, indicating that they correspond to ``true"
eigenvalues of the full (infinite-dimensional) problem.
The other two branches vary significantly with $M$.
Although not visible with this resolution, all eigenvalues have a strictly negative
real part, meaning that the asynchronous state is stable.
A clearer view of the spurious exponents is presented in Fig.~\ref{f4}(b), where the eigenvalues are suitably rescaled.
The good overlap indicates that the real parts increase linearly with $M$, while the imaginary components decrease as
$1/M^{3/2}$. Altogether, this means that the corresponding directions are increasingly stable and therefore harmless
in dynamical simulations.
The change of stability can be appreciated in Fig.~\ref{f4}(c), where we plot the relevant part of the spectrum for
three different $K$ values: 80 (red circles), 160 (blue crosses), and 1280 (green triangles).
In the last case, a pair of complex-conjugate eigenvalues has crossed the $y$ axis, indicating the occurrence of a
Hopf bifurcation. Here, performing quasi-adiabatic simulations of the FPE by varying $K$, we have verified that
the transition is super-critical.

\begin{figure}[h!]
\includegraphics*[width=0.45\textwidth]{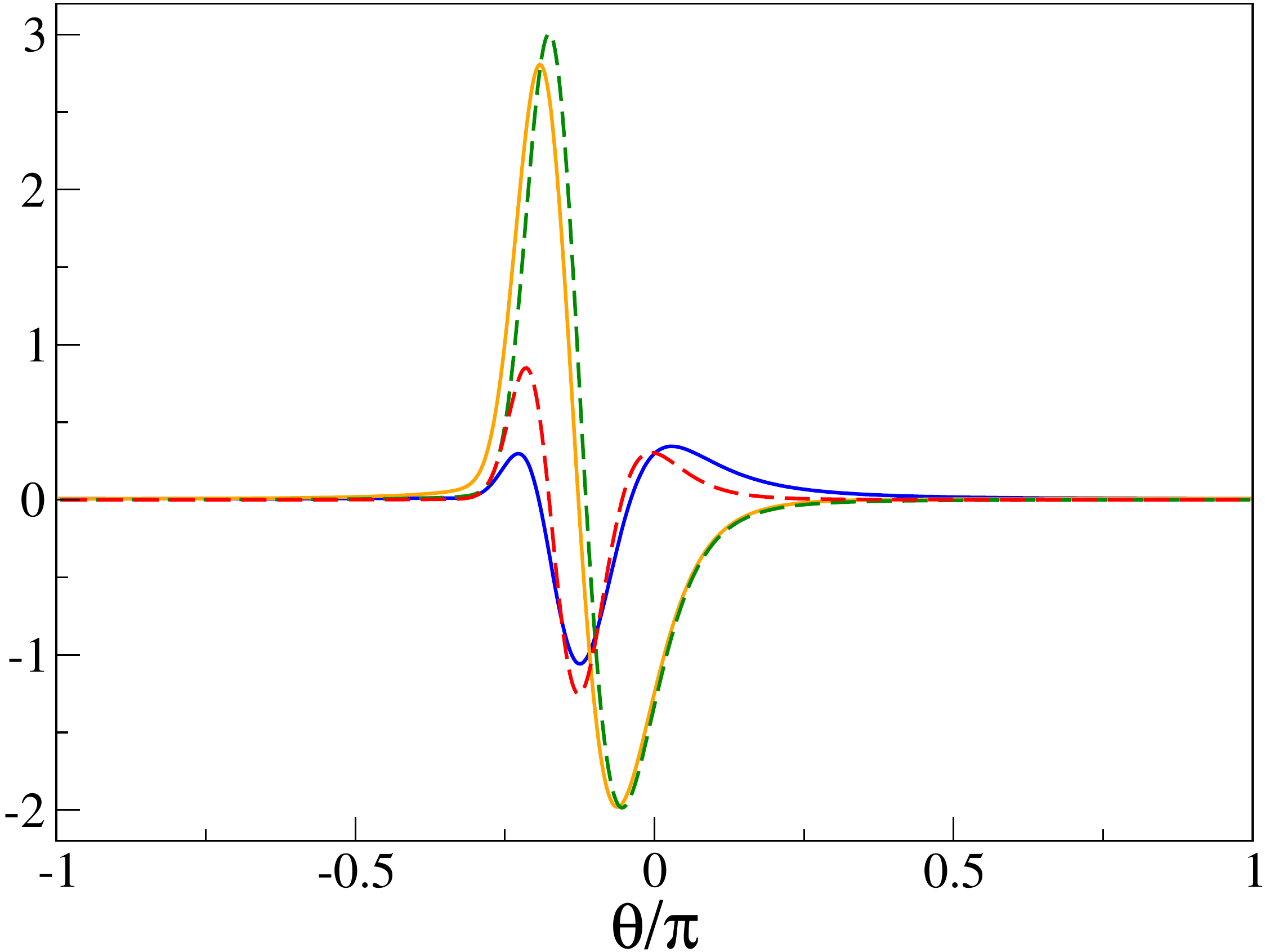}
\caption{Relevant perturbations of the stationary PDF $R^{(0)}(\theta)$,
which contribute to its periodic oscillations (orange and blue solid lines).
The green dashed line refers to $d R^{(0)}(\theta) / d \theta$
and the red dashed line to  $d^2 R^{(0)}(\theta) / d \theta^2$
reported in arbitrary units. Parameters as in Fig. \ref{f4}.
}
\label{f5}
\end{figure}

The linear stability analysis also allows determining the eigenvectors.
In particular, it is instructive to estimate the first two
${\bf e}_1$ and ${\bf e}_2$, as they identify the manifold over which the periodic oscillations unwind.
Since they are complex conjugated, it suffices to focus on the real and imaginary parts separately.
Via inverse Fourier transform (in order to obtain the representation in $\theta$-space) we obtain the
two functions reported in Fig.~\ref{f5} (see the orange and blue solid lines). As expected, since they can be seen as perturbations
of the probability density, they have zero average. Moreover, they closely resemble
the first and second derivative of the PDF $R_0(\theta)$ (see the green and red dashed lines).

\subsubsection{2CCs analysis}

Here, we discuss the linear stability of the asynchronous state with reference to the
2CCs approximation. The evolution equation in tangent space is obtained by linearizing Eqs.~\eqref{2CC},
\begin{widetext}
\begin{eqnarray}
&& \delta \dot{z}_1 = \delta z_1 (i A_{g_0} +i) + i z_1^{(0)} \delta A_{g_0} +
\delta H (1 + \kappa^{(0)}_2 + (z^{(0)}_1)^2 ) + H^{(0)}(\delta \kappa_2 + 2 z^{(0)}_1 \delta z_1) \nonumber
\\
&& - \delta D_{g_0} \left[\frac{(1+z^{(0)}_1)^3}{2}  \right]
- \frac{3}{2} D_{g_0} \left[ \delta z_1 (1+  z_1^{(0)})^2   \right]
\\
&& \delta \dot{\kappa}_2 = 2(i A_{g_0} + i) \delta \kappa_2 + 2 i \delta A_{g_0} \kappa^{(0)}_2
+ 4 \delta H z_1 \kappa_2 + 4 H^{(0)} ( \delta z_1 \kappa^{(0)}_2 + z_1^{(0)} \delta \kappa_2 )
\nonumber
\\
&& - \delta D_{g_0} \left[ \frac{1}{2} (1+z^{(0)}_1)^4 +6 \kappa^{(0)}_2 (1+z^{(0)}_1)^2  \right]
-  D_{g_0} \left[2 (1 +  z^{(0)}_1) \delta z_1 \left((1+z_1^{(0)})^2 +6 \kappa_2 \right)
+6 \delta \kappa_2 (1+z^{(0)}_1)^2
  \right]
\end{eqnarray}
\end{widetext}
where
$$
H^{(0)}=\frac{1}{2}\left[i(A_{g_0}-1)\right] \; ,
$$
Upon then differentiating the definitions of $A_g$, $D_g$ and $H$, we obtain
\begin{eqnarray}
& & \delta  A_{g_0} = - \sqrt{K} g_0 \delta \nu \; , \; \delta D_{g_0} = \frac{g_0^2}{2} \delta \nu
\; , \; \delta H = i\delta A_g/2 \\
& & \delta \nu = \frac{1}{\pi} Re \left\{\left[ -\frac{\delta z_1}{(1+ z_1^{(0)})^2} + \frac{\delta \kappa_2}{(1+z^{(0)}_1)^3}
- \frac{ \kappa^{(0)}_2  \delta z_1 }{(1+z^{(0)}_1)^4} \right]  \right\}
\nonumber
\label{lincoef_CC}
\end{eqnarray}
so that the model is complete.

For $i_0 = 0.006$ and $g_0=1$, a Hopf bifurcation is detected for $K^{HB} \simeq   54$.
The bifurcation is subcritical, meaning that the oscillations persist also below $K^{(c)}$, actually until $K^{SN} \simeq 35$ where
they disappear via a saddle-node bifurcation of limit cycles. In the following section this scenario is compared with the other approaches.

\subsection{Emergence of Collective Oscillations}

\begin{figure}
\centerline{\includegraphics[width=0.45\textwidth]{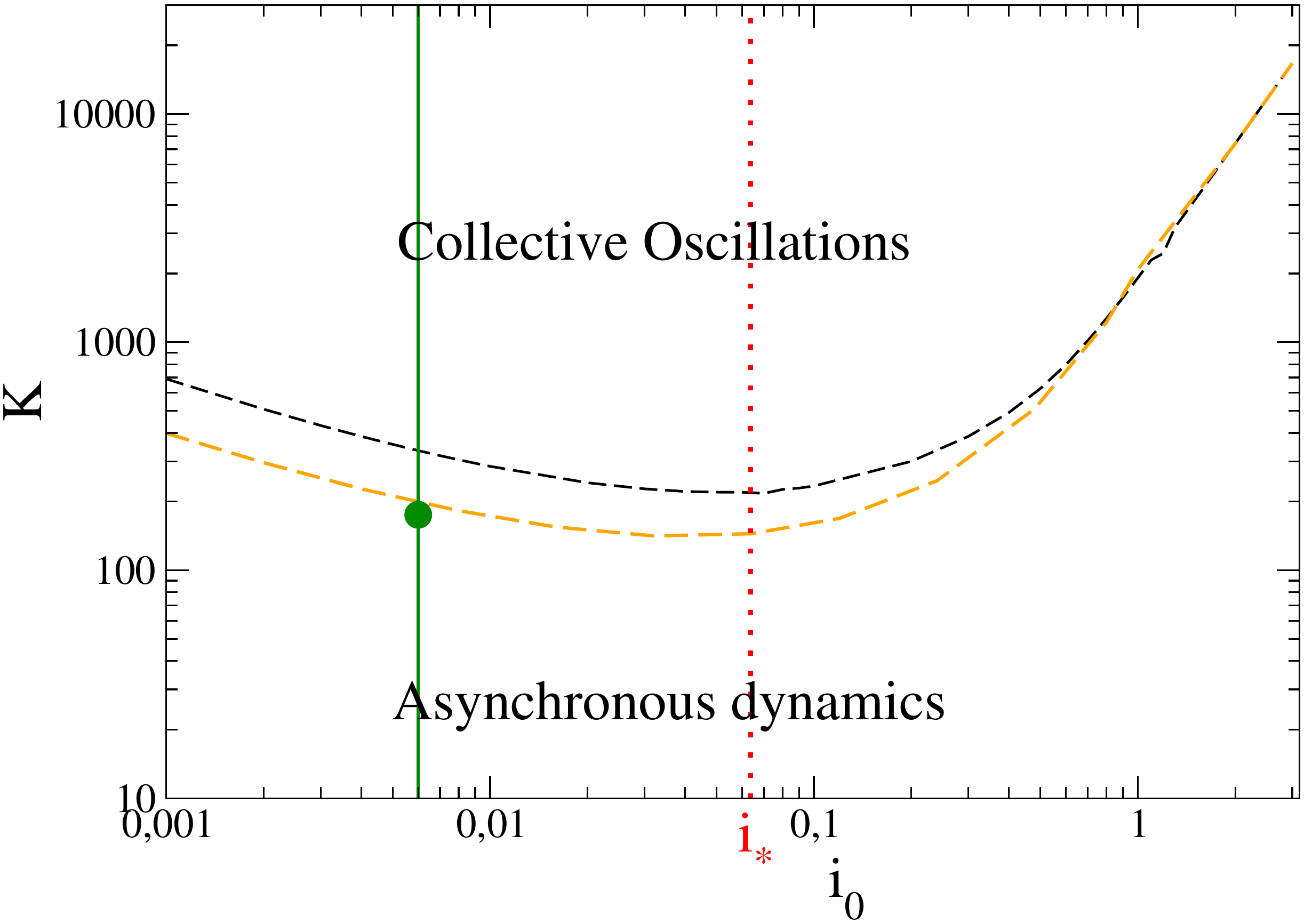}}
\caption{Phase diagram as a function of the mean connectivity $K$ and the DC current $i_0$. Black (orange) dashed line show the super-critical Hopf bifurcation line estimated from the stability analyses of the fixed point in the Fokker Planck model with $M=128$ modes, under the Poissonian (renewal with $CV=0.8$) approximation. The red dashed vertical line corresponds to the critical current $i_*$
\eqref{eq:202} separating the fluctuation from the mean driven regimes, as reported in Fig.~\ref{fig2} (a). The green vertical line refers to the parameter cut analysed in Fig.~\ref{f6} while the green dot corresponds to the transition point observed in direct
numerical simulations (for more details see Fig. \ref{f6}).
Other parameters: $g_0=1$.}
\label{phase_d}
\end{figure}

As mentioned in the previous sub-section, the instability of the
asynchronous state leads to periodic COs via a super-critical Hopf bifurcation.
In Fig.~\ref{phase_d} we report in the plane $(i_0,K)$ the transition lines separating the asynchronous
states from COs obtained within the Poissonian approximation (black dashed line)
and the renewal approximation with $CV=0.8$ (orange dashed line). 
For sufficiently large $i_0 > i_*$, when the dynamics is balanced but mean-driven, 
the two curves coincide and the statistics of the spike trains seem to be irrelevant.
However, for low currents the transition occurs for Poissonian statistics at larger $K$ 
with respect to the renewal approximation. This is consistent with the fact that the spike trains
are more irregular in the Poissonian case and therefore the collective effects emerge
for larger $K$.

\begin{figure}
\centerline{\includegraphics[width=0.45\textwidth]{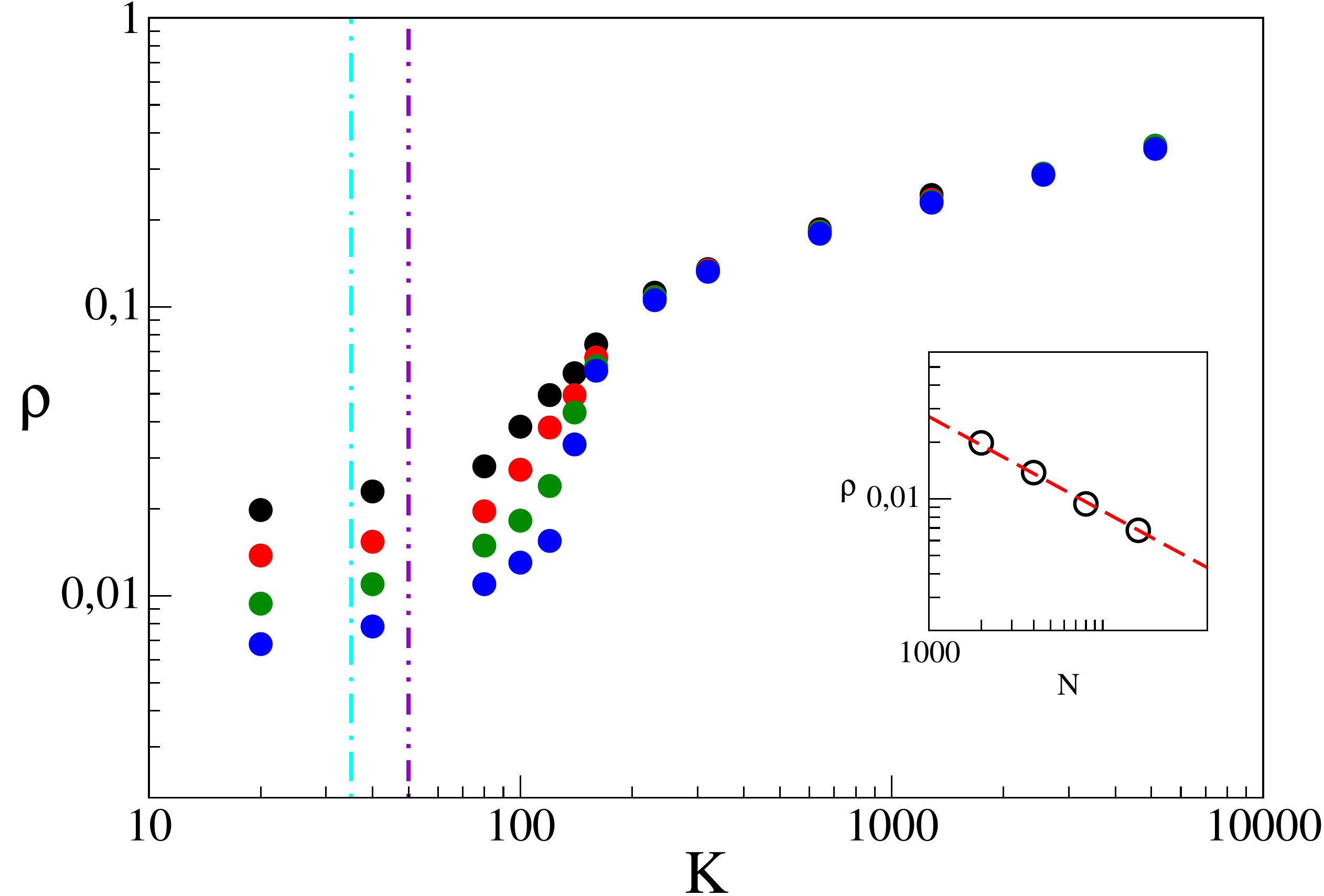}}
\caption{Order parameter $\rho$ versus the in-degree K for different
network sizes: $N=2000$ (black circles), 4000 (red circles),
8000 (green circles) and 16000 (blue circles).
The two-dots-dashed vertical violet (dot-dashed cyan)
line indicates $K^{HB}$ ($K^{SN}$) for the sub-critical Hopf  (saddle node) bifurcation point obtained within the 2CCs approximation.
The inset report the scaling of $\rho$ versus $N$ for $K=20$, the red
dashed line corresponds to a power law $N^{-1/2}$. Parameters as in Fig. \ref{f4}.
}
\label{f6}
\end{figure}

We have directly explored the behavior of the network for
the specific value of $i_0=0.006 < i_*$. The transition point $K^{(c)} \simeq 170 - 180$ 
(see the green dot in Fig.~\ref{phase_d}), is very close to the theoretical renewal prediction,
while the Poisson approximation is significantly larger.
More detailed results are reported in Fig.~\ref{f6}, where we plot $\rho$ (see Eq.~\eqref{rho}) for increasing in-degrees $K$
and different system sizes, namely $N=2000$, 4000, 8000 and 16000.

For completeness, also the prediction of the 2CCs approximation is reported (dot-dashed magenta line).
In this case, as already discussed, the Hopf bifurcation is sub-critical: periodic oscillations appear for $K=K_{SN}^0 \simeq 35$,
before the asynchronous state loses stability at $K=K_{HB}^{0}=50$.
The agreement with the direct numerical simulations is definitely worse, but one
should not forget that this is a low-dimensional model in a context (homogeneous network),
where the OA is not attractive.

\begin{figure}
\centerline{\includegraphics[width=0.45\textwidth]{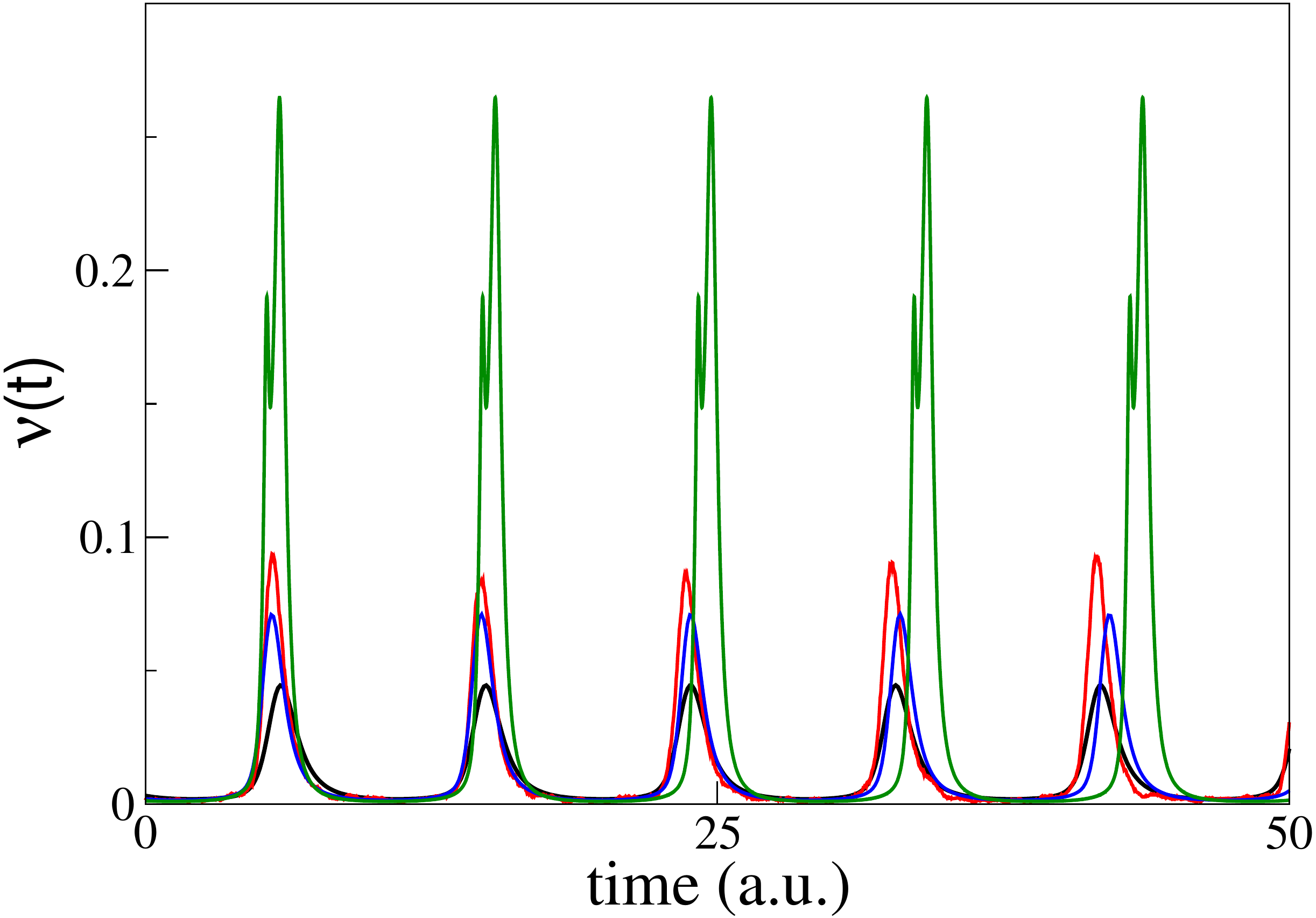}}
\caption{Instantaneous firing rate $r(t)$ versus time. The data refer to
network simulations with $N=16000$ (red line), to MF solutions obtained
by truncating the FPE to $M=64$ modes for the Poissonian noise (black line)
or the Renewal approximation with $CV=0.8$ (blue line), as well as
to the 2CCs approximation (green line). Parameters as in Fig. \ref{f4} and $K=640$.
}
\label{f7}
\end{figure}

Let us now analyze the COs.
In Fig.~\ref{f7} we report the instantaneous firing rate for $K=640$.
The FPE with Poissonian noise nicely reproduces the period
of the oscillations, although their amplitude
is substantially underestimated.
The renewal approach with $CV=0.8$ ensures a better representation of the
oscillation amplitude, but the period is slightly longer
(blue line in Fig.~\ref{f7}). Finally the 2CCs approximation overestimates
both the amplitude and the period of the COs (see the green line).

Finally, we analyze the scaling behavior of the oscillatory dynamics in the limit of large median in-degree $K$.
Roughly speaking, the frequency $\nu_{CO}$ of the collective oscillations increases with $K$, as well as the average firing rate
$\nu$, suggesting increasing deviations from the balanced regime.
In fact, we also see that the instantaneous firing rate oscillates between a maximum, which increases with $K$ and a minimum, which
decreases, while, simultaneously the width of the peaks shrinks (with reference to Fig.~\ref{f7}, the peaks become taller and thinner, when
$K$ is increased.).

\begin{figure}
\centerline{\includegraphics[width=0.48\textwidth]{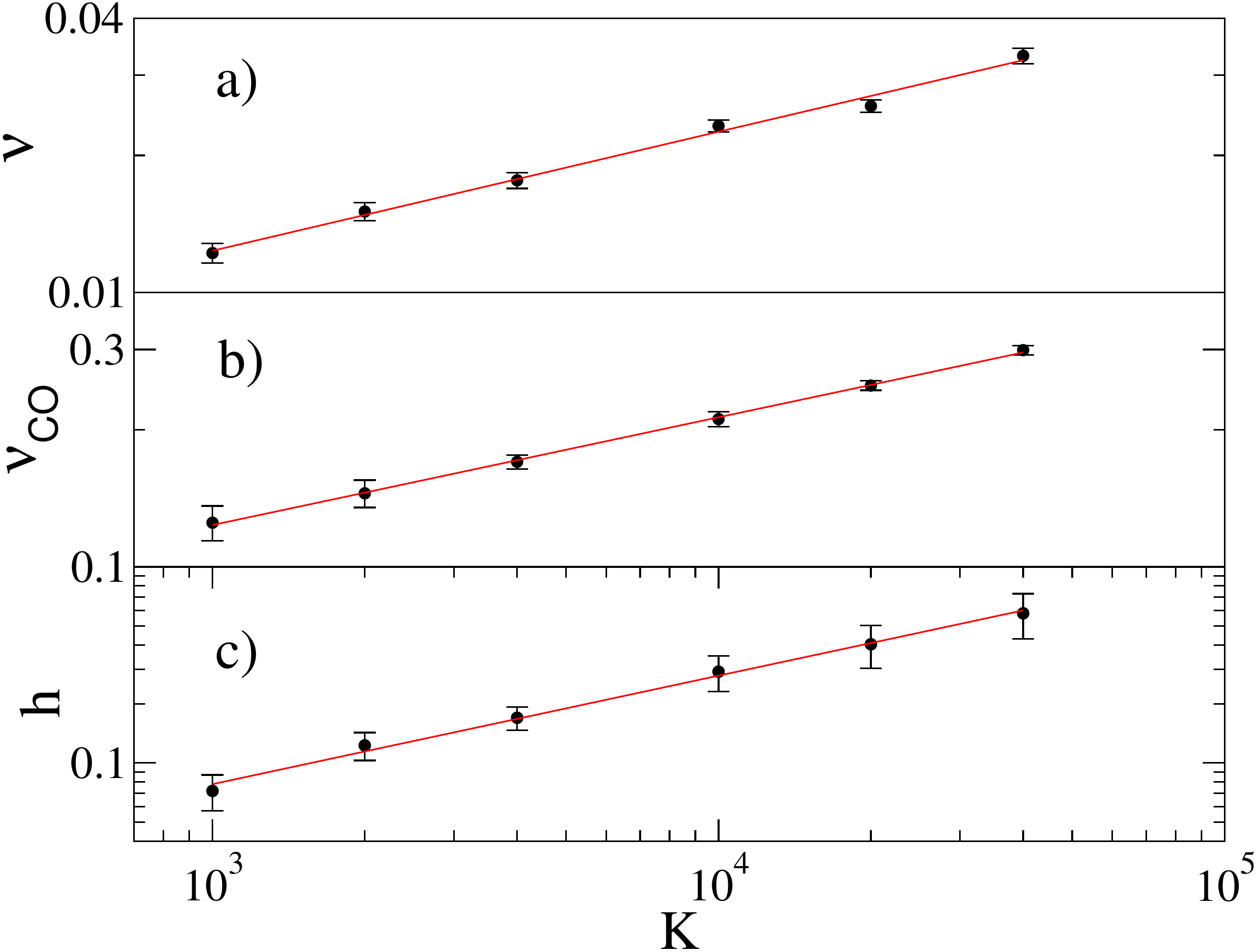}}
\caption{Average (over time) firing rates $\nu$ (panel a), frequencies of the COs $\nu_{CO}$ (panel b) and heights $h$ of the peak of COs (panel c)
versus the in-degree $K$. Data are obtained by integrating the annealed mean field model (for the annealed model see sub-section III D)) with $N_{ann}=10^4$ neurons. Error bars are estimated as the standard deviation over 100 cycles of the COs. The red line shows the fit with a power $K^{\alpha}$,  $\alpha =0.26$ in panel a), $\alpha =0.24$ in panel b) and $\alpha =0.55$  in panel c). Other parameters are $i_0=0.006$, $g_0=1$ and $\Delta_g=0$.
}
\label{fscal}
\end{figure}

The results of a quantitative analysis are reported in Fig.~\ref{fscal}(a). They have been obtained by simulating an annealed network of
$10000$ neurons. We have preferred to simulate a network, rather than integrating the FPE, because numerical instabilities 
make it difficult to perform reliable simulations for large $K$.
A power-law fit of the data in panel (b) suggests that $\nu_{CO} \approx K^{0.24}$, very close to the scaling behavior
\begin{equation}
\nu_{CO} \simeq  K^{1/4}  \; .
\label{scaling_nu}
\end{equation}
predicted by the MF model~\eqref{montbrio} for the frequency of damped oscillations around the stable MF focus. 

More intriguing is the scaling behavior of the firing rate, $\nu \approx K^{0.26}$, since it basically coincides with the 
maximum possible rate reachable in absence of inhibition.
In fact, upon neglecting inhibition, the membrane potential dynamics is ruled by the equation
\begin{equation}
\dot V = I + V^2 = i_0 \sqrt{K} + V^2 \; .
\end{equation}
Upon rescaling $V$ as $U=V/K^{1/4}$ and time as $\tau = tK^{1/4}$, the differential equation rewrites as
\begin{equation}
\dot U = i_0 + U^2 \; .
\end{equation}
Since the time (in $\tau$ units) for $U$ to travel from $-\infty$ to $+\infty$ is of order $\mathcal{O}(1)$, the ISI in
the original time frame is $\mathcal{O}(K^{-1/4})$, which obviously represents
a lower bound for the average ISI. Remarkably, the inhibition, unavoidably induced by the synaptic coupling, does not 
alter significantly the scaling of the average firing rate with $K$.

\begin{figure}
\centerline{\includegraphics[width=0.48\textwidth]{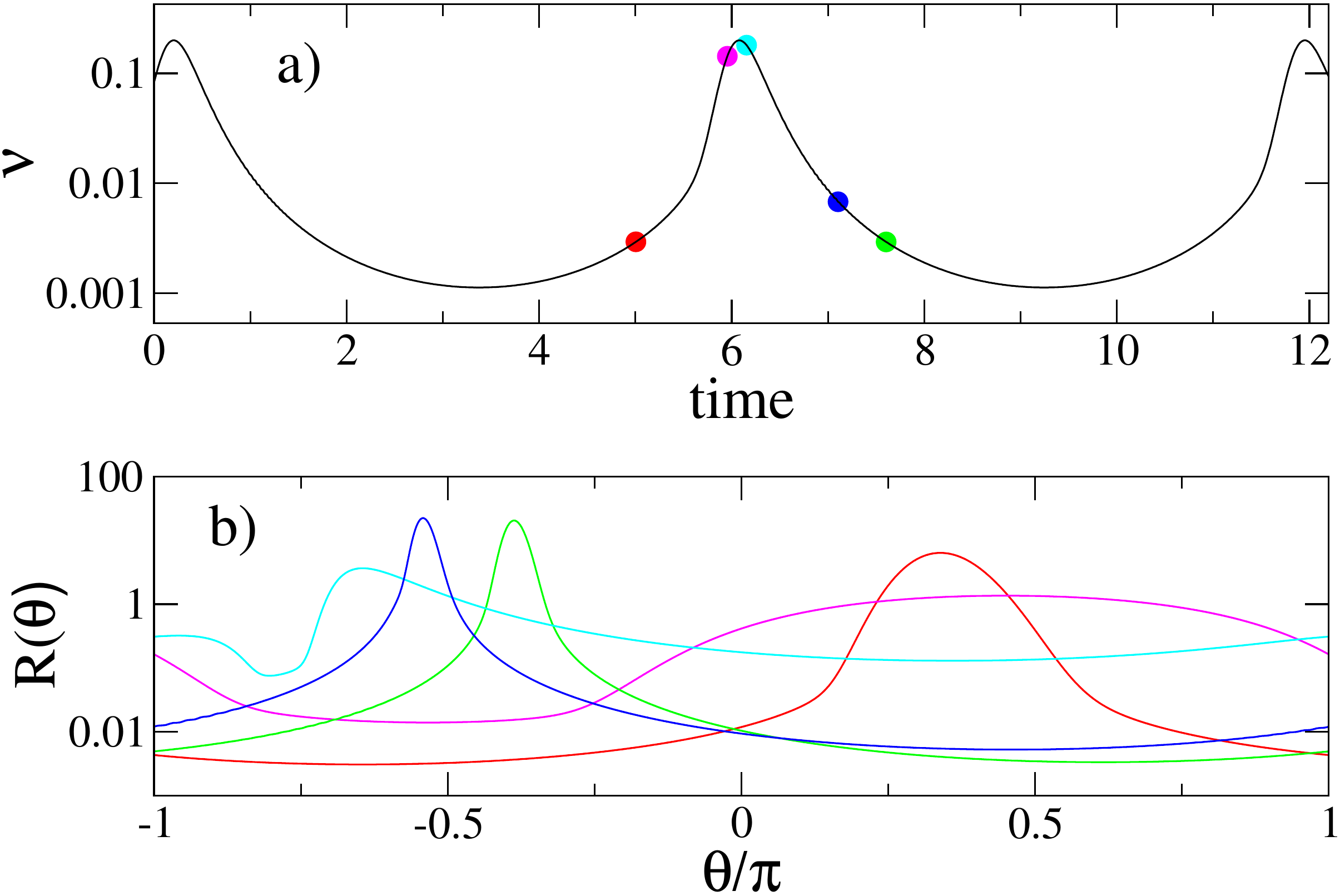}}
\caption{In the top panel we show the  the firing rate $\nu$ in time estimated from the FPE truncated at $M=128$ by estimating the current fluctuations within the Poisson approximation Eq. \eqref{poisson}. In the lower panel we report the PDFs $R{(\theta)}$ versus the angle $\theta$ at different times (see the corresponding dots in the top panel). Parameters: $i_0 = 0.006$, $g_0=1$, $K=4000$ and $\Delta_g=0$.}
\label{prob_time}
\end{figure}

Numerical simulations suggest that the temporal profile of $\nu(t)$ is significantly different
from zero only during tiny time intervals of duration $\Delta t \approx K^{-\alpha}$, separated by a time interval $T = 1/\nu_{CO}$
(see panel (a) in Fig.~\ref{prob_time}, where the simulations have been performed by integrating the FPE for $K=4000$ -
please notice the logarithmic vertical scale).
By assuming that the height of the peaks scales as $h \approx K^{\beta}$, it follows that the average firing rate scales as
\begin{equation}
\nu \approx  \frac{\Delta t}{T}h \approx K^{\beta-\alpha+1/4}
\end{equation}
where we have inserted the known scaling behavior of the collective oscillations.
Thus, we see that $\nu$ can scale as $\nu_{CO}$ provided that $\alpha=\beta$ or, equivalently, that the number of neurons 
which emit a spike in a single burst is independent of $K$, which is precisely the behavior observed in the numerical
simulations.
Notice, that the periodic behavior of the collective dynamics does not imply that each neuron fires periodically.
As testified by the large CV value (namely,  $CV \simeq 0.8$), the neural activity keeps being irregular in the limit of large $K$.

Next, we discuss the value of $\beta$. At the time of the maximum rate, $K^\beta$,
 the (inhibitory) current received by each neuron is $I_c \approx K^{\beta+1/2}$, 
much larger than the excitatory external current of order $K^{1/2}$, which can then be neglected.
In such conditions each neuron sees a potential $-V^3/3 + K^{\beta+1/2} V$, where the second term follows from the inhibitory
coupling. Hence, we are in the presence of a deep minimum of the effective potential located 
in $V_{min} \approx -K^{\beta/2+1/4}$ and a maximum in $V_{max} \approx +K^{\beta/2+1/4}$. 
All $V$ values smaller than $V_{max}$ are attracted towards the minimum and do not contribute to the ongoing burst.
Those above the maximum, instead, will unavoidably reach the threshold.
The time needed for nearly all of such neurons to fire is about half of the burst width $\Delta t$ and can be obtained by
integrating the evolution equation
\begin{equation}
\dot V = V^2 - K^{\beta+1/2}
\end{equation}
from an initial condition slightly larger than $V_{max}$ up to infinity. It is easily seen that $\Delta t \approx K^{-\beta/2-1/4}$.
It follows that the area of the peak is
\begin{equation}
A = h \delta T \approx K^{\beta/2-1/4} \; .
\end{equation}
Having numerically evidence that $A$ is independent of $K$, it finally follows that $\beta=1/2$.
This prediction is consistent with the numerical observations reported in
Fig.~\ref{fscal}(c), where a fit of the numerical data yields 0.55. Given the presence of statistical fluctuations
(the peak height fluctuates because of the finiteness of the number of neurons)  and
the probable presence of deviations due to subleading terms, the agreement is satisfactory.

We conclude this section with some considerations on the nature of the oscillatory regime arising in the limit of large $K$.
In Fig.~\ref{prob_time}(a), we report the temporal profile of the firing rate $\nu$ for a not too large $K$-value. The vertical logarithmic
scale indicates that the neural activity oscillates between almost silent intervals and short bursts characterized by a  strong activity.
This might suggest a nearly synchronous regime, but this is not the case.
In panel (b) of the same figure, we plot five snapshots of  
the distribution of the $\theta$ angles (this is preferable to the $V$ representation, as the $\theta$ values are bounded).

There we see that just before the burst, the distribution is very broad (please notice the vertical logarithmic scale).
Then it is strongly narrowed during the peak (as a consequence of the self-built strongly confining potential mentioned above).
At the same time, the peak is first pushed backward (so long as inhibition is strong) and then starts drifting forward and,
simultaneously broadens.
Altogether, the manifestation of a narrow peak of the neural activity is the consequence of an increasingly fast
dynamics due to the fact that many membrane potentials find themselves in a region where their ``velocity" is very large.

\section{Heterogeneous case}

In this Section we consider the heterogeneous case, assuming that the in-degrees $k$ are 
Lorentzian distributed~\eqref{eq:2} with median $K$ and HWHM
$\Delta_k = \Delta_0 \sqrt{K}$. As already discussed in Section III,
while introducing the Langevin approach, the in-degree disorder can be treated as quenched disorder of
the effective synaptic couplings $g$ -- also Lorentzian distributed. 
As in the homogeneous case, the Langevin formulation can be mapped onto a
FPE (see Eq.~\eqref{FP_ZM}) for the Kuramoto-Daido order parameters, and one eventually can get rid of the disorder by
invoking the Chaucy's residue theorem.

In principle, one derive an expression for the average firing-rate by integrating
the analytic expression (\ref{eq:103}) of the firing rate $\nu_g$ of each specific
sub-population over the distribution of the synaptic couplings
\begin{equation}
\nu=\int_{-\infty}^{+\infty} L(g)\,\nu_g\mathrm{d}g \quad .
\label{eq:301}
\end{equation}
However, as explained in Appendix A, despite the distribution being Lorentzian, we cannot 
derive in this case an analytic expression.
This is because of essential singularities within
the integration contour, which prevent the application of the residue theorem.

\begin{figure}
\centerline{\includegraphics[width=0.45\textwidth]{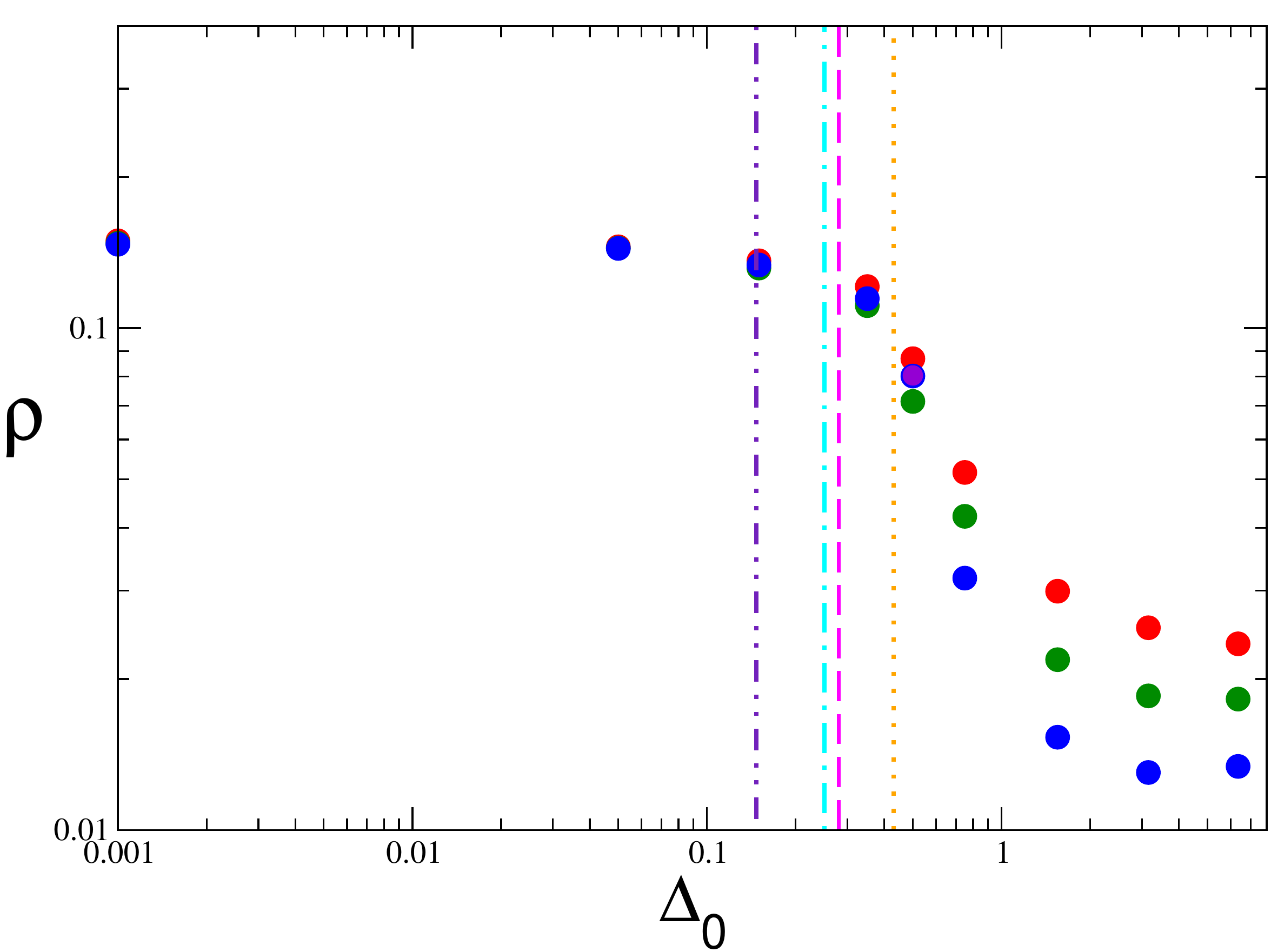}}
\caption{Heterogeneous model : order parameter $\rho$ versus $\Delta_0$.
Symbols refer to direct simulations of the network for different system sizes:
$N=4000$ (red), 8000 (green), 16000 (blue) and 32000 (violet).  The vertical magenta dashed (orange dotted) line denotes $\Delta_{0}^{HB}$ corresponding to the super-critical Hopf bifurcation
identified from the analysis of the FPE truncated
to $M=64$ within a Poissonian (renewal) approximation where the amplitude of the current fluctuations is given by Eq. \eqref{poisson} (Eq.\eqref{renewal} with $CV=0.8$). The
vertical violet two-dots-dashed (cyan dot-dashed) line indicates $\Delta_{0}^{HB}$ ($\Delta_{0}^{SN}$) for the sub-critical Hopf  (saddle node) bifurcation point as obtained within the 2CCs approximation. The data have been also averaged over 20 different network realizations. Other parameters: $K=400$, $i_0 = 0.006$, $g_0 =1$.
}
\label{f9}
\end{figure}

The analysis presented in the previous section has shown that COs arise in homogeneous networks
for sufficiently large median in-degrees $K$ and small external currents $i_0$. 
By continuity, it is reasonable to conjecture that the same occurs in networks with moderate heterogeneity
(this regime has been indeed reported in Ref.~\cite{matteo}).
In Fig.~\ref{f9} we plot the order parameter $\rho$ versus the parameter controlling 
the structural disorder $\Delta_0$ (see its definition below Eq.~\eqref{eq:2}),
for different network sizes $N$.
There, we see that the COs observed in the homogeneous case $\Delta_0 =0$  persist
up to a critical value $\Delta_0^{(c)} \simeq 0.40$, when the structural disorder becomes so large as to wash out collective phenomena.
Indeed, above $\Delta_0^{(c)}$  $\rho$ scales as $1/N^{1/2}$ as expected for
asynchronous dynamics. To better understand the 
transition, let us recall that in \cite{matteo}, the authors noticed that 
the the average coefficient of variation $CV$
displays a finite value $CV \simeq 0.8$ in the region where COs are observable, while it vanishes 
above $\Delta_0^{(c)}$. This was explained by conjecturing that for increasing $\Delta_0$
only few neurons, the ones with in degrees proximal to the median $K$, can
balance their activity, while the remaining neurons are no longer able
to satisfy the balance conditions, as recently shown in Refs.~\cite{landau2016,pyle2016}.

The bifurcation diagram is well reproduced by the linear stability analysis of the FPE,
which predicts a super-critical Hopf bifurcation at  $\Delta_{0}^{HB} \simeq 0.28$
($\Delta_{0}^{HB} \simeq 0.43$) in the Poissonian (renewal) approximation: 
see the magenta dashed line (orange dotted line) in Fig.~\ref{f9}. 
The linear stability analysis of the 2CCs approximation instead predicts a sub-critical Hopf bifurcation,
accompanied, as usual, by a coexistence interval $[0.16,0.24]$. 
$\Delta_0^{HB} \simeq 0.24$ (see the double-dotted-dashed violet line) is the critical point, where
the asynchronous regime loses stability, while
$\Delta_0^{SN} \simeq 0.16$ (see the dot-dashed, cyan line) corresponds to the 
saddle-node bifurcation where stable COs coalesce with analogous unstable oscillations.
Since direct numerical simulations and the FPE
do not show any evidence of a bistable region close to the critical point, it follows that this
bistability is a spurious effect of the 2CCs approximation.

\begin{figure}
\centerline{\includegraphics[width=0.45\textwidth]{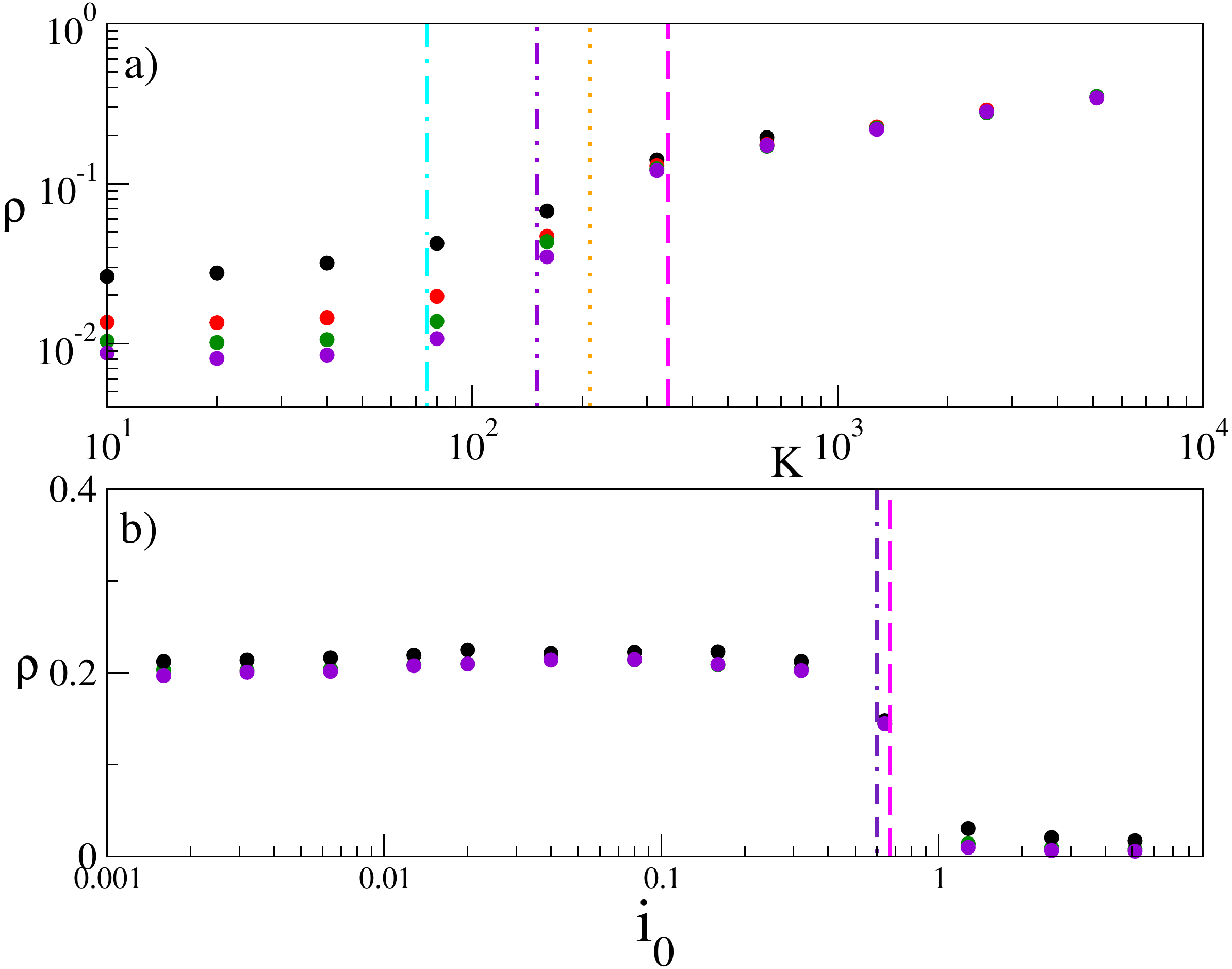}}
\caption{Transitions for the heterogeneous model : order parameter $\rho$ versus $K$ (panel a) and versus $i_0$ (panel b). Symbols refer to direct simulations of the network for different system sizes: $N=8000$ (red), 8000 (green), 16000 (blue) and 32000 (violet).  In panel (a) the vertical magenta dashed (orange dotted) line denotes $K^{HB}$ as estimated within a MF approach for the FPE truncated to $M=64$ with a Poissonian (renewal) approximation where the amplitude of the current fluctuations are given by Eq.~\eqref{poisson} 
(Eq.~\eqref{renewal} with $CV=0.8$). The
vertical violet two-dots-dashed (cyan dot-dashed)
line indicates $K^{HB}$ ($K^{SN}$) for the sub-critical Hopf  (saddle node) bifurcation point obtained within the 2CCs approximation. In panel (b) the vertical  violet dot-dashed (magenta dashed) line indicates the critical value of $i_{0}^{HB}$ at which there is a super-critical Hopf bifurcation as obtained with the 2CCs approximation (with the FPE truncated
to $M=64$ within a Poissonian approximation).
Parameters: $K=1000$, $i_0 = 0.006$, $g_0 =1$ and $\Delta_0 = 0.1$, when not differently specified.
}
\label{f10}
\end{figure}

The robustness of COs has been studied also by decreasing the in-degree $K$ and increasing the input current $i_0$. 
The results are shown in Fig.~\ref{f10} for $\Delta_0 =0.1 < \Delta_0^{(c)}$.
In panel (a), the current is set equal to $i_0 = 0.006$. Numerical simulations indicate a transition to the asynchronous regime at
$K^{(c)} \simeq 200-250$. The scenario is well captured by the Fokker-Planck analysis which predicts 
a super-critical Hopf bifurcation at $K^{HB} \simeq 343$ ($K^{HB} \simeq 210$) within the Poissonian (renewal) approximation.
Also in this case, the 2CCs model predicts a sub-critical Hopf bifurcation at $K^{HB} \simeq 75$ (cyan dot-dashed line) 
accompanied by a saddle-node bifurcation of the limit cycles at $K^{SN} \simeq 150$ (magenta dot-dashed line).
Fig.~\ref{f10}(b) refers to $K=1000$. In this case, direct numerical simulations, the FPE, and the 2CCs approximation, all 
predict a super-critical bifurcation around $i_0^{(c)} \simeq 0.6-0.7$.

\begin{figure}
\centerline{\includegraphics[width=0.45\textwidth]{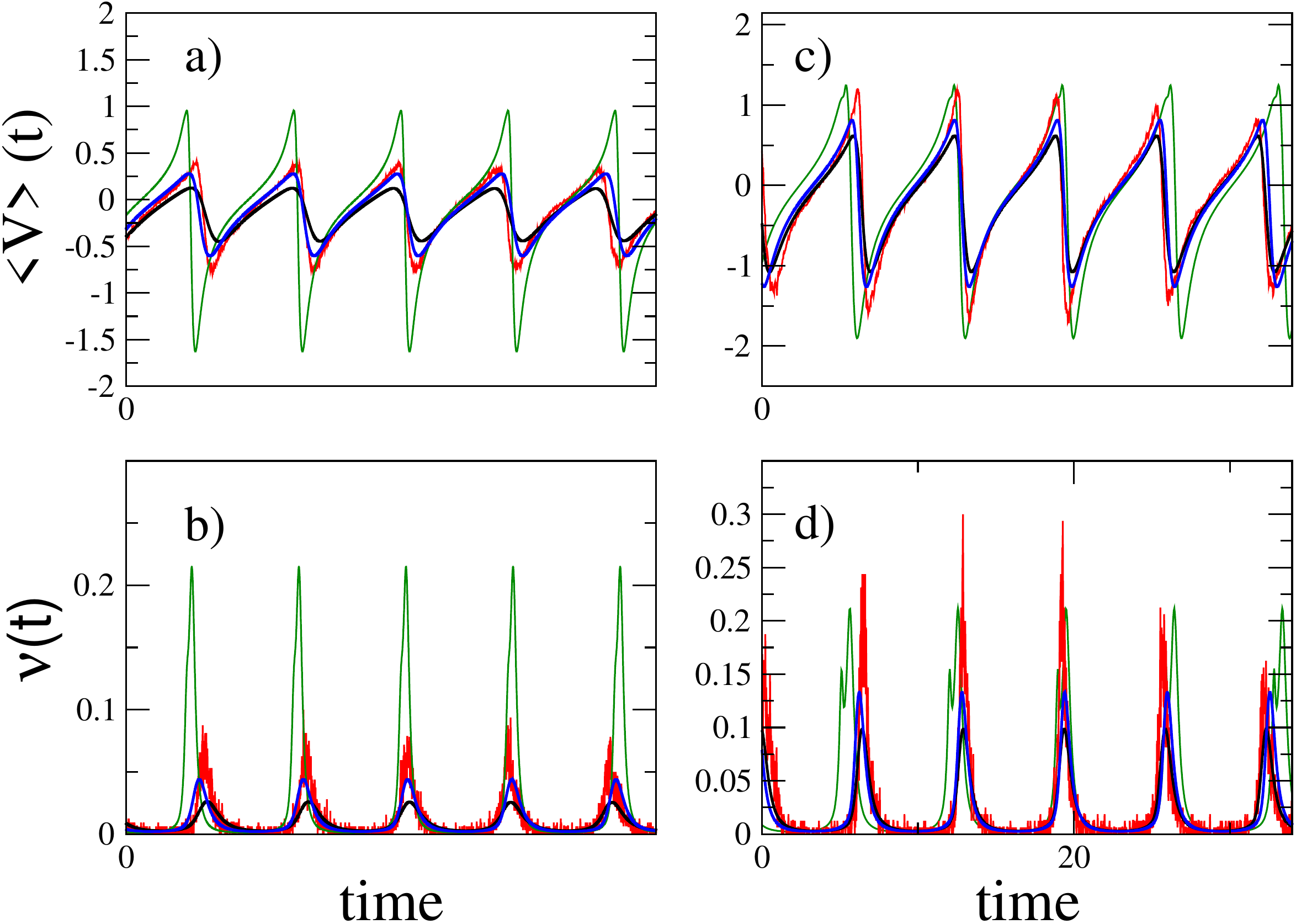}}
\caption{Time trace of the mean membrane potential $v(t)$ (panels (a,c)) and of the firing rate $\nu$
(panels (b,d)).
The parameters $(K,i_0)$ are $(500,0.006)$ in (a,b) and  $(1000,0.01)$  in (c,d).
The red lines always refer to direct simulations for $N=16000$ (red line).
Black (blue) lines correspond to the integration of the FPE, truncated
after $M=64$ Fourier modes for a Poissonian noise (renewal noise with $CV=0.8$).
Finally green lines correspond to the 2CC approximation. The structural heterogeneity is $\Delta_0 = 0.1$.
}
\label{ftraces}
\end{figure}

Finally, in Fig.~\ref{ftraces} we report the evolution of the mean membrane potential $v(t)$ and of the population firing rate
$\nu(t)$ for two different sets of parameter values. 
We compare the results of network simulations (red solid lines) with the outcome of the FPE 
in the Poissonian (black solid line) and renewal (blue solid line) approximation,
as well as with the behavior of the 2CCs model (green solid lines).
The agreement between direct simulations and the results of the FPE with renewal
noise are remarkable, including the shape of the oscillations.
The 2CCs approximation works better than in the homogeneous case, but it still
overestimates the amplitudes of the COs and slightly the period.

On the one hand, we can conclude that the FPE reproduces the dynamics of heterogeneous networks with a good quantitative accuracy.
On the the other hand the 2CCs, while being able to capture the transition from COs to the 
asynchronous regime, is much less precise both in terms of oscillations shape and the nature of the transition.

\section{Conclusions}

This article has been devoted to a mean field characterization of sparse balanced networks composed of identical QIF neurons 
both with homogeneous and heterogeneous in-degree distributions. 
The main focus of our analysis has been the spontaneous emergence of coherent or collective fluctuations
out of the asynchronous balanced regime. 
Collective oscillations are the result of an internal macroscopic coherence and may, in general, be either 
regular or irregular~\cite{luccioli2010, olmi2010, olmi2011}.
They resemble coherent fluctuations observed across spatial scales in the neocortex~\cite{srinivasan2007,volgushev2011,okun2012}.
In the present setup COs are strictly periodic and arise even for completely homogeneous in-degrees.

Somehow similarly to what previously done for Integrate-and-Fire neurons
\cite{brunel1999,brunel2000,mattia2002} the starting point is the formulation of 
a Langevin equation for the membrane potential, where the noise is self-consistently
determined by assuming that the fluctuations of the input current follows from the
superposition of independent stochastic processes: the single-neuron spiking trains.
Two main assumptions are made while formulating the Langevin description: Poisson and renewal statistics.
The in-degree heterogeneity has a twofold effect:
it acts as a quenched disorder in the synaptic couplings and as
an additional parameter affecting the noise amplitude.

The Langevin equations are turned into a family of FPEs for the
evolution of the distributions of the membrane potentials for each sub-population characterized by a given in-degree.
The Fokker-Planck formulation is twice infinite dimensional: 
as it deals with the distribution of membrane potentials and for its dependence on the in-degree connectivity.
The latter dependence can be removed by assuming a Lorentzian distribution of the in-degrees,
in which case the evolution equation reduces to a single FPE which depends on the median in-degree 
and on a parameter controlling the width of the structural heterogeneity of the distribution,
(similarly to what done in \cite{ratas2019} for a globally coupled QIF network subject to external noise terms). 

Altogether, the FPE proves very accurate both in the description of homogeneous and heterogeneous networks.
The renewal approximation is typically more precise than the Poisson approximation.
However, stronger deviations are expected for very small currents and not-too-large connectivity.
In such conditions, the ``granularity" of the input signal received by every neuron cannot be
anymore neglected and the white noise assumption underlying the FPE should be 
replaced by shot noise as already done for Leaky Integrate-and-Fire neurons in \cite{richardson2010,olmi2017}.
Future studies will be devoted to this specific aspect.

A further simplification is then proposed and explored, by expanding
the probability distribution of membrane potentials into circular cumulants~\cite{tyulkina2018}.
The fast (exponential) decrease of the cumulant amplitude with their order suggests truncating the hierarchy 
after two cumulants. The quality of the 2CCs approximation has been tested both for the description of
the asynchronous regime and the onset of COs. 
Interestingly, the 2CCs approximation works reasonably well also in homogeneous networks
where the Ott-Antonsen manifold is not attractive.
Nonetheless, in some cases the 2CCs model reproduces incorrectly the nature of the Hopf bifurcation 
(sub- instead of super-critical); moreover, the amplitude of the oscillations is substantially larger than
in real networks. 
Anyway the value of the 2CCs model 
relies on its low-dimensionality: 
it should be appreciated that two variables are able to captures the onset of COs via a Hopf bifurcation and predict  
reasonable values for the stationary firing rate when the asynchronous regime is stable.
In a future perspective, possible improvements should be explored.
In particular, the inclusion of the third cumulant, although this
issue requires an in-depth analysis: as the amplitude
of the cumulants decreases very rapidly with their order, it is unclear why a third cumulant should play
a relevant role.

The balanced regime has been invoked as  mechanism explaining irregular low firing activity in the cortex. 
It is commonly believed that in a balanced asynchronous regime the system operates sub-threshold,
where the activity is driven by current fluctuations \cite{bal1,brunel2000}. However, as shown
in \cite{lerchner2006} this is not the only possible scenario: 
both mean- and fluctuation-driven balanced asynchronous regimes can emerge 
in an excitatory-inhibitory network dominated by the inhibition drive for finite $K$. 
Our analysis confirms that both regimes can emerge in a fully
inhibitory network for arbitrarily large connectivity.
In particular, fluctuation (mean) driven balanced dynamics appear for 
small (large) DC currents as well as for large (small) inhibitory synaptic coupling.
Furthermore, we have also shown that  a perfectly balanced
regime can be obtained by fine tuning of the parameters for any finite in-degree.

For what concerns the regime characterized by the presence of collective oscillations,
the large $K$-limit proves very interesting since the dynamics exhibits increasingly
strong deviations from a balanced regime. First of all the frequency of the
collective oscillations diverges as $\nu_{CO} \propto K^{1/4}$, as
also suggested by the linear stability analysis of the MF solution.
Remarkably, the average firing rate scales in the same way: this is due
to the occurrence of the concentration of the activity in short but very
strong bursts.
With the help of numerical observations showing that the percentage of neurons
participating to the population bursts is independent by $K$, we have concluded that the height 
of the bursts grows as $K^{1/2}$.  It would be desirable to draw this
conclusion in a more rigorous way.

In the heterogeneous case, we examined three different scenarios for the emergence of COs:
namely, COs can arise at large $K$, as well as for sufficiently low
structural heterogeneity $\Delta_0$ and input currents $i_0$. 
All these transitions are captured both from the Fokker-Planck formulation as well as
from the 2CCs approximation, this at variance with the low dimensional MF formulation reported in \cite{matteo} 
that was based on the Ott-Antonsen Ansatz and therefore not including the current fluctuations. These results clearly indicate that the
role of coherent fluctuations present in the balanced regime is fundamental for the birth of COs. Therefore the neurons should be
in the fluctuation driven regime, usually observable at low $i_0$, and their dynamics should be sufficiently coherent to promote oscillations at the network level, as it occurs for low $\Delta_0$ and large $K$.

\acknowledgments{We acknowledge extremely useful discussions with  L. Klimenko , G. Mongillo, S. Olmi, and E. Shklyaeva. 
AT received financial support by the Excellence Initiative I-Site Paris Seine (Grant No ANR-16-IDEX-008), by the Labex MME-DII (Grant No ANR-11-LBX-0023-01) and by the ANR Project ERMUNDY (Grant No ANR-18-CE37-0014) (together with MdV), all part of the French programme ``Investissements d'Avenir''. The derivation and study of the exact solution for the firing rate were supported by the Russian Science Foundation (Grant No. 19-42-04120).
}

\appendix

\section*{Appendix A}

In this Appendix, we will demonstrate that the integral \eqref{eq:301}
cannot be performed via the  residue theorem, as usually expected, due
to the presence of essential singularities within
the integration contour.

Let us  clarify the properties of the analytic function $\nu_g = D_g^{1/3} \mathcal{R}(\xi)$ (\ref{eq:103}),
which is expressed in terms of $n$-th order Bessel functions $J_n$ and $I_n$ of the first kind and modified, respectively.
Therefore, we should first recall the definition and some properties of the Bessel functions:
\begin{equation}
\mathrm{I}_\alpha(Z)=i^{-\alpha}\mathrm{J}_\alpha(iZ) \equiv\sum_{m=0}^{\infty}\frac{\left(\frac{Z}{2}\right)^{2m+\alpha}}{m!\,\Gamma(m+\alpha+1)}\,,
\label{eq:302}
\end{equation}
which possesses the property:
\begin{equation}
\mathrm{I}_\alpha(e^{in\pi}Z)=e^{i\alpha n\pi}\mathrm{I}_\alpha(Z)\,,
\label{eq:303}
\end{equation}
where $n$ is an integer. Hence,
\begin{equation}
\mathrm{I}_{\pm\frac13}(Z)
 =\mathrm{I}_{\pm\frac13}\big(|Z|e^{i(n\pi+\theta)}\big)
 =e^{\pm i\frac{n\pi}{3}}\mathrm{I}_{\pm\frac13}\big(|Z|e^{i\theta}\big)\,,
\label{eq:304}
\end{equation}
where $\theta\in(-\pi/2;\pi/2]$ (for the convenience of computer calculations).

By employing Eqs.~(\ref{eq:302}) and (\ref{eq:304}), one finds that
\begin{align}
&\mathrm{I}_{\pm\frac13}(y^{3/2})=\mathrm{I}_{\pm\frac13}\big((-e^{i\pi}y)^{3/2}\big)
 =\mathrm{I}_{\pm\frac13}(e^{i2\pi}e^{-i\frac{\pi}{2}}(-y)^{3/2})
\nonumber\\
&
\quad
=e^{\pm i\frac{2\pi}{3}}e^{\mp i\frac{\pi}{6}}\mathrm{J}_{\pm\frac13}\big((-y)^{3/2}\big)
 =e^{\pm i\frac{\pi}{2}}\mathrm{J}_{\pm\frac13}\big((-y)^{3/2}\big)\,.
\nonumber
\end{align}
This result tells us that the expression of $\nu_g$ for $A_g <0$ (first line in \eqref{eq:103})
can be obtained from the expression for $A_g >0 $ (second line in \eqref{eq:103}) simply by setting
$A_g=e^{i\pi}(-A_g)$. Therefore, for the moment we can limit our analysis to the case $A_g < 0$.

\begin{figure}[!t]
\centerline{
\includegraphics[width=0.49\textwidth]%
 {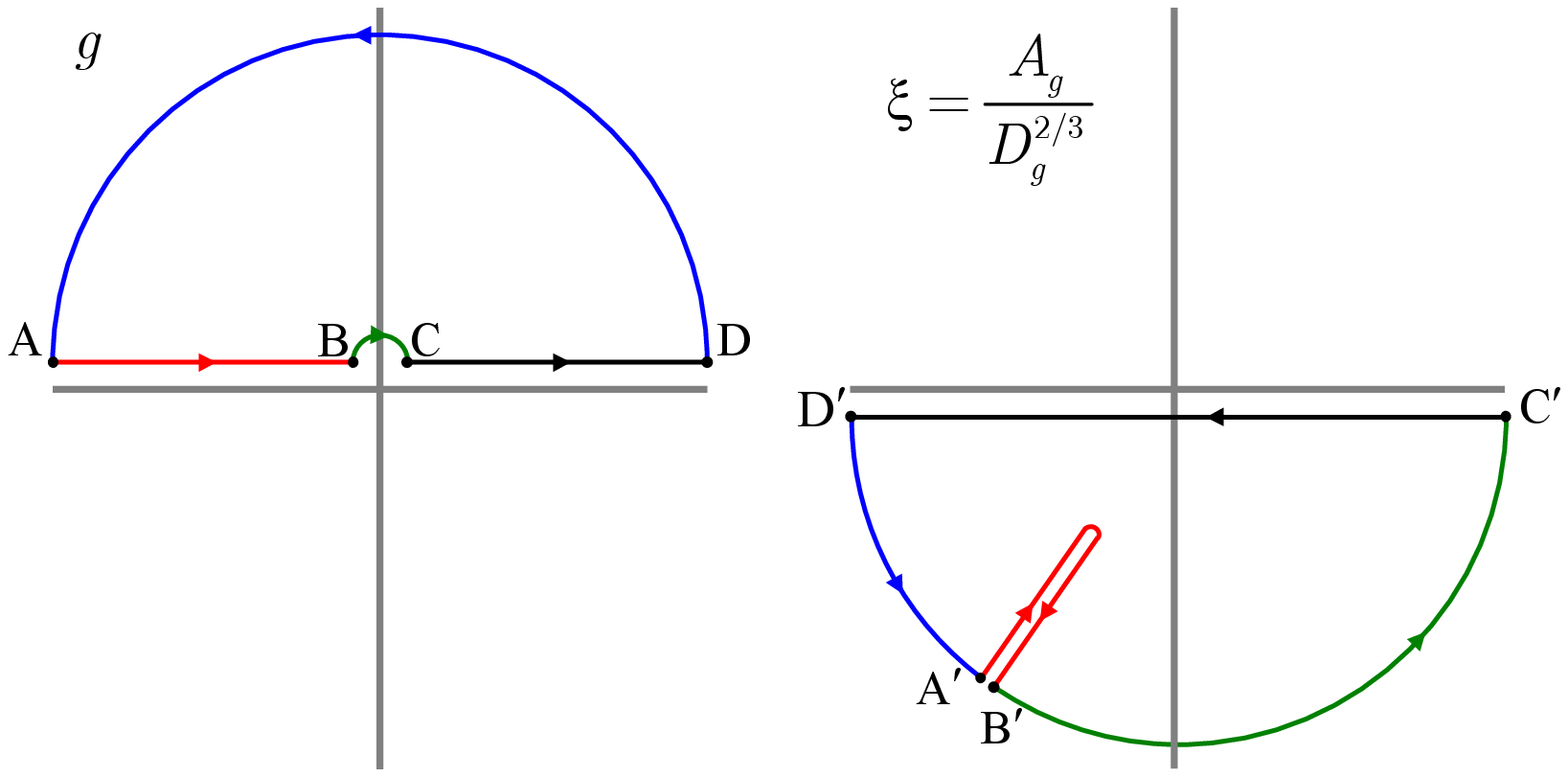}
}
\caption{The closed integration contour (right) on the complex $\xi$-plane corresponds to the one on the complex $g$-plane (left). The points $\mathrm{A}^\prime$, $\mathrm{B}^\prime$, $\mathrm{C}^\prime$, $\mathrm{D}^\prime$ correspond to $\mathrm{A}$, $\mathrm{B}$, $\mathrm{C}$, $\mathrm{D}$.
}
  \label{fig3}
\end{figure}

Since $\nu_g = D_g^{1/3} \mathcal{R}(\xi)$, where $\xi = A_g/D_g^{2/3}$,
in order to estimate the integral~\eqref{eq:301}
we should define a closed integration contour in the complex $\xi$-plane.
This contour is shown in Fig.~\ref{fig3}.
To apply the residue theorem we should identify the poles
of the function $\mathcal{R}(\xi)\equiv \mathcal{R}([3/2]^{2/3}y)$
with
\[
y\equiv\left(\frac{2}{3D_g}\right)^\frac23 A_g\;,
\]
where the new variable $y$ is introduced for the brevity of calculations.
We can now rewrite the expression appearing in the denominator in Eq.~(\ref{eq:103}) for $A_g >0$ as:
\begin{widetext}
\begin{align}
&\big[\mathrm{J}_{\frac13}(y^{3/2})\big]^2 +\big[\mathrm{J}_{-\frac13}(y^{3/2})\big]^2 -\mathrm{J}_{\frac13}(y^{3/2})\,\mathrm{J}_{-\frac13}(y^{3/2})
\nonumber\\
&=\mathrm{J}_{\frac13}(y^{3/2})\,e^{-i\frac{\pi}{3}}\mathrm{J}_{\frac13}(e^{i\pi}y^{3/2}) +\mathrm{J}_{-\frac13}(y^{3/2})\,e^{i\frac{\pi}{3}}\mathrm{J}_{-\frac13}(e^{i\pi}y^{3/2})
-\mathrm{J}_{\frac13}(y^{3/2})\,\mathrm{J}_{-\frac13}(e^{i\pi}y^{3/2}) -\mathrm{J}_{\frac13}(e^{i\pi}y^{3/2})\,\mathrm{J}_{-\frac13}(y^{3/2})
\nonumber\\[5pt]
&=\left[e^{-i\frac{\pi}{6}}\mathrm{J}_{\frac13}\big(y^{3/2}\big) -e^{i\frac{\pi}{6}}\mathrm{J}_{-\frac13}\big(y^{3/2}\big)\right]
 \left[e^{-i\frac{\pi}{6}}\mathrm{J}_{\frac13}\big((e^{i\frac{2\pi}{3}}y)^{3/2}\big) -e^{i\frac{\pi}{6}}\mathrm{J}_{-\frac13}\big((e^{i\frac{2\pi}{3}}y)^{3/2}\big)\right]
\nonumber\\[5pt]
&=\left[\mathrm{J}_{\frac13}\big((e^{i\frac{2\pi}{3}}y)^{3/2}\big) +\mathrm{J}_{-\frac13}\big((e^{i\frac{2\pi}{3}}y)^{3/2}\big)\right]
 \left[\mathrm{J}_{\frac13}\big((e^{-i\frac{2\pi}{3}}y)^{3/2}\big) +\mathrm{J}_{-\frac13}\big((e^{-i\frac{2\pi}{3}}y)^{3/2}\big)\right].
\label{eq:305}
\end{align}
The expression for the subpopulation firing rate $\nu_g$ can be rewritten as:
\begin{align}
\nu_g= D_g^{1/3} \mathcal{R}(\xi) = D_g^{1/3} \mathcal{R}([3/2]^{2/3}y) =
\frac{\frac{9 D_g}{4\pi^2 A_g}} {
\left[\mathrm{J}_{\frac13}\big((e^{i\frac{2\pi}{3}}y)^{3/2}\big) +\mathrm{J}_{-\frac13}\big((e^{i\frac{2\pi}{3}}y)^{3/2}\big)\right]
 \left[\mathrm{J}_{\frac13}\big((e^{-i\frac{2\pi}{3}}y)^{3/2}\big) +\mathrm{J}_{-\frac13}\big((e^{-i\frac{2\pi}{3}}y)^{3/2}\big)\right]}\;,
\end{align}
\end{widetext}
which yields (\ref{eq:103}) for positive and negative $A_g$.

Thus, the function $\mathcal{R}(\xi)$ possesses the two sets of poles with $\mathrm{arg}(\xi)=\pm2\pi/3+2\pi n$, since a multivalent analytic function $[\mathrm{J}_{1/l}(z)\pm\mathrm{J}_{-1/l}(z)] =\sum_{m=0}^{\infty}C_{1/l,m}z^{2m+1/l} \pm\sum_{m=0}^{\infty}C_{-1/l,m}z^{2m-1/l}$ with positive integer $l$ and real-valued coefficients $C_{\alpha,m}$ possesses zeros only for $z=\rho$ and $z=e^{il\pi}\rho$, where $\rho$ is real positive~\cite{properties}.
The sequences of these poles form an essential singularities at $\xi=|\infty|e^{i(2\pi n\pm\pi/3)}$ and the integration path runs through it (as we show below). Hence, the integration contour cannot be closed via infinity at the upper/lower half-plane of $\xi$ (or $y=(2/3)^{2/3} \xi$), and the residue theorem cannot be employed.

It is now important to find the path on the complex $\xi$-plane corresponding to $g$ varying from $-\infty$ to $+\infty$:
\begin{align}
\xi&=\frac{A_g}{D_g^{2/3}}=\frac{\sqrt{K}(i_0 - g \nu)}{\left(\frac{g_0 g \nu}{2}\right)^{2/3}} \;.
\label{eq:306}
\end{align}
On the segment $\mathrm{C}\mathrm{D}$ (see Fig.~\ref{fig3}): $g>0$; therefore, $\xi$ is real and runs from $+\infty$ to $-\infty$. On the arc $\breve{\mathrm{B}\mathrm{C}}$: $\xi=e^{-i\frac{2}{3}\arg(g)}\sqrt{K}i_0/[(g_0 |g| \nu)/2]^{2/3}$, which is an arc at infinity (as $|g|\to0$), running from $|\infty|e^{-i2\pi/3}$ to $+|\infty|$ (see the arc $\breve{\mathrm{C}^\prime\mathrm{D}^\prime}$ in Fig.~\ref{fig3}). On $\mathrm{A}\mathrm{B}$: $g=e^{i\pi}|g|$ and $\xi=e^{-i2\pi/3}\sqrt{K}(i_0+|g|\nu)/[(g_0|g|\nu)/2]^{2/3}$ forms the arc $\breve{\mathrm{A}^\prime\mathrm{B}^\prime}$. On $\breve{\mathrm{D}\mathrm{A}}$: $g=|\infty|e^{i\alpha}$ and $\xi=\sqrt{K}(2/g_0)^{2/3}e^{i\pi}(g\nu)^{1/3}$, which is the arc $\breve{\mathrm{D}^\prime\mathrm{A}^\prime}$.
The segment $\breve{\mathrm{A}^\prime\mathrm{B}^\prime}$ passes exactly along the line with the poles, $\mathrm{arg}(x)=-2\pi/3$. Not only the mutual position of these poles and the integration path needs to be clarified for finite $x$ (the left and right segments can be on the one side of poles or `envelope' them), but, more importantly, at infinity we have an essential singularity at $\xi=|\infty|e^{-i2\pi/3}$ and the contribution of its vicinity into the integral is uncertain. Thus, as announced above, the residue theorem cannot be employed for this case.

\bibliographystyle{apsrev4-1}

\bibliography{lif_ost_cd}

\begin{thebibliography}{59}
\bibitem [{con()}]{cont}%
  \BibitemOpen
  \href@noop {} {}\bibinfo {note} {In the majority of programming languages and
  packages for analytical calculations, complex-valued functions $f(z)$ are
  defined with $\mathrm{arg}(z)\in(-\pi;\pi]$, while in our case one must
  consider the Bessel functions $\mathrm{I}_{\pm1/3}(z)$ for
  $\mathrm{arg}(z)\in[0;\pi)$. Thus, a blind calculations of expression for
  $A_g < 0$ by employing a standard continuation software almost certainly will
  yield incorrect results for $A_g > 0$, while correct explicit calculations
  for $(-A_g)=A_ge^{i\pi}$ yield the equation in the second line
  of~(\ref{eq:103}). One also should bear in mind that $\mathrm{I}_{\pm1/3}(z)$
  is a three-sheet analytic function, while $\mathrm{I}_{\pm1/3}(x^{3/2})$ is a
  two-sheet one.}\BibitemShut {Stop}%
\bibitem [{pro()}]{properties}%
  \BibitemOpen
  \href@noop {} {}\bibinfo {note} {One should check two properties here. First,
  for $z=\rho e^{i\varphi}$, one finds $\mathrm{J}_{1/l}(\rho e^{i\varphi})
  \pm\mathrm{J}_{-1/l}(\rho e^{i\varphi})
  =e^{i\varphi/l}\sum_{m=0}^{\infty}C_{1/l,m}\rho^{2m+1/l}e^{i2m\varphi} \pm
  e^{-i\varphi/l}\sum_{m=0}^{\infty}C_{-1/l,m}\rho^{2m-1/l}e^{i2m\varphi}$. If
  $\varphi=n\pi$ with integer $n$, then $e^{2m\varphi}=1$ and the sums yield
  the Bessel functions of the first kind with series of zeros; if
  $\varphi=(n+1/2)\pi$, then $e^{2m\varphi}=(-1)^m$ and the sums yield the
  Bessel functions of the second kind with no zeros; for any other value of
  $\varphi$ the sums yield a complex-valued function of real-valued $\rho$ and
  generally will have no zeros. However, for $\varphi=n\pi$,
  $[\mathrm{J}_{1/l}(z)\pm\mathrm{J}_{-1/l}(z)]
  =e^{i\varphi/l}\mathrm{J}_{1/l}(\rho)\pm
  e^{-i\varphi/l}\mathrm{J}_{-1/l}(\rho)$ and these two terms can sum-up to
  zero only if $\varphi/l-(-\varphi/l)=j\pi$ with integer $j$. Note, whilst for
  integer $\alpha$, function
  $\mathrm{J}_{-\alpha}(\rho)=(-1)^\alpha\mathrm{J}_{\alpha}(\rho)$, for
  noninteger $\alpha$ (in our case $\alpha=1/3$) functions
  $\mathrm{J}_{-\alpha}(\rho)$ and $\mathrm{J}_{\alpha}(\rho)$ are independent
  and the degenerate case of coinciding zeros of
  $\mathrm{J}_{\alpha}(\rho_\ast)=0$ and $\mathrm{J}_{-\alpha}(\rho_\ast)=0$ is
  not possible. Combining conditions $\varphi=jl\pi/2$ and $\varphi=n\pi$, for
  odd $l$, we find that the function of our interest can have zeros only for
  $\varphi=ln\pi=3\pi n$. Second, the crossings of the abscissa axis by the
  function $\mathrm{J}_{1/3}(x^{3/2}) \pm\mathrm{J}_{-1/3}(x^{3/2})$ are all
  transversal, i.e., the function possesses only first order
  zeros.}\BibitemShut {Stop}%
\end{thebibliography}

\end{document}